\documentclass[aps,prx,twocolumn,superscriptaddress,longbibliography]{revtex4-1}
\usepackage{amsmath}
\usepackage{amssymb,amsfonts}
\usepackage{amsmath,amsthm,amssymb,bbold}
\usepackage{graphicx,float,caption,subcaption}
\usepackage{tikz}

\usetikzlibrary{math}
\usepackage[colorlinks]{hyperref}

\newcommand{\figref}[1]{Fig.~\ref{#1}}
\newcommand{\appref}[1]{Appendix~\ref{#1}}
\newcommand{\tabref}[1]{Table~\ref{#1}}
\newcommand{\ket}[1]{|#1\rangle}

\newcommand{\be}{\begin{equation}}
\newcommand{\ee}{\end{equation}}

\newcommand{\opi}{O_{i}^{ab}}

\newcommand{\op}[3]{O_{#1}^{#2#3}}

\def\bea{\begin{eqnarray}}
\def\eea{\end{eqnarray}}

\def\c{\chi}

\def\s{\sigma}

\def\Tr{\mathrm{Tr}}

\newcommand{\T}[2]{C^{#1}_{#2}}
\newcommand{\pp}{\psi}
\newcommand{\si}{\sigma}
\newcommand{\I}{\mathbb{1}}

\newcommand{\dO}[1]{O^{(-#1_l, \cdots,-#1_2,-#1_1)}}
\newcommand{\ldO}[2]{O^{(-#2_{l+1},-#1_l, \cdots,-#1_2,-#1_1)}}

\newcommand{\HH}[2]{H^{#1_l, \cdots,#1_2,#1_1}_{#2_l, \cdots,#2_2,#2_1}}
\newcommand{\lHH}[2]{H^{#1_{l+1},#1_l,\cdots,#1_2,#1_1}_{#2_{l+1},#2_l,\cdots,#2_2,#2_1}}
\newcommand{\ldHH}[2]{H^{#1_{l+1},#1_l,\cdots,#1_2,#1_1}_{#2_l,\cdots,#2_2,#2_1}}
\newcommand{\CC}[2]{C^{#1_l, \cdots,#1_2,#1_1}_{#2_l, \cdots,#2_2,#2_1}}
\newcommand{\lCC}[2]{C^{#1_{l+1},#1_l,\cdots,#1_2,#1_1}_{#2_{l+1},#2_l,\cdots,#2_2,#2_1}}
\newcommand{\ldCC}[2]{C^{#1_{l+1},#1_l,\cdots,#1_2,#1_1}_{#2_l,\cdots,#2_2,#2_1}}
\newcommand{\lsCC}[3]{C^{#1_{l+1},#1_l,\cdots,#1_2,#1_1}_{#3_{l+1},#2_l,\cdots,#2_2,#2_1}}
\newcommand{\ind}[1]{\{#1\}}
\newcommand{\dd}[1]{\frac{d#1}{2\pi i}}
\newcommand{\vO}[2]{O_{#1}^{(-\vec{#2})}}
\newcommand{\vC}[2]{C^{\vec{#1}}_{\vec{#2}}}

\begin{document}

\title{ Precision reconstruction of rational CFT from exact fixed point tensor network}
\author{Gong Cheng}
\thanks{These authors contribute equally.}
\affiliation{Department of Physics, Virginia Tech, Blacksburg, VA 24060, USA}
\affiliation{Maryland Center for Fundamental Physics, University of Maryland, College Park, MD 20740, USA}
\author{Lin Chen}
\thanks{These authors contribute equally.}
\affiliation{School of Physics and Optoelectronics, South China University
of Technology, Guangzhou 510641, China}
\author{Zheng-Cheng Gu}
\email{corresponding to: zcgu@phy.cuhk.edu.hk}
\affiliation{Department of Physics, The Chinese University of Hong Kong, Shatin, New Territories, Hong Kong, China}
\author{Ling-Yan Hung}
\email{corresponding to: lyhung@tsinghua.edu.cn}
\affiliation{Yau Mathematical Sciences Center, Tsinghua University, Haidian, Beijing 100084, China}
 


\begin{abstract}
The novel concept of entanglement renormalization and its corresponding tensor network renormalization technique have been highly successful in developing a controlled real space renormalization group (RG) scheme.
Numerically approximate fixed-point (FP) tensors are widely used to extract the conformal data of the underlying conformal field theory (CFT) describing critical phenomena.
In this paper, we present an explicit analytical construction of the FP tensor for 2D rational CFT. We define it as a correlation function between the "boundary-changing operators" (BCO) on triangles. Our construction fully captures all the real-space RG conditions.
We also provide concrete examples, such as Ising, Yang-Lee and Tri-critical Ising models to compute the scaling dimensions explicitly based on the corresponding FP tensor.  The BCO descendants turn out to be an optimal basis such that truncation in bond dimensions naturally produces comparable accuracies with the leading existing FP algorithms.
Interestingly, our construction of FP tensors is closely related to a strange correlator, where the holographic picture naturally emerges. Our results also open a new door towards understanding CFT in higher dimensions. 

\end{abstract}
\maketitle


\section{Introduction}  
In the past two decades, the novel concept of entanglement renormalization \cite{VidalER, Gu08, Vidal09a, Vidal09b, Gu09, PhysRevLett.102.180406} has been developed to study 
critical systems. In particular, computationally efficient algorithms has been proposed based on tensor network techniques, such as various schemes of tensor network renormalization (TNR)  \cite{Levin,Gu08,Gu09,Xie, HOTRG, VidalTNR, Yang, PhysRevB.95.045117}.  
It is found that even with a moderate size of bond dimensions kept in the coarse graining procedure, there are lots of important information such as central charge, scaling dimensions and operator product expansion (OPE) coefficient of conformal field theory (CFT) can be read off from the fixed point (FP) tensors, which are approximate fixed points of the TNR algorithms \cite{Gu09,Yang}. 
Furthermore, TNR algorithms have been applied to study CFT topological defects \cite{PhysRevB.94.115125} and conformal boundaries \cite{PhysRevB.100.035449, PhysRevB.82.161107}. 

Despite the huge successes in numerically extracting conformal data through tensor network simulations, the analytical construction of FP tensors for critical systems remains a significant challenge. While progress has been made in understanding the components of FP tensors associated with primary fields 
 \cite{PhysRevResearch.4.023159,UedaFP}, generalizing these constructions for descendant fields remains unclear. On the other hand, the recently proposed holographic picture and generalized symmetry description \cite{Ji:2019jhk, Kong:2020cie, catc} for CFT suggest that the complete algebraic structure of FP tensors might provide us an alternative way to understand CFT. It suggests that fundamentally a continuous field theory admits a rigorous discrete formulation, and such a novel formulation is a revolution in modern physics with far-reaching consequences, such as systematically simulating interacting field theories numerically, which is otherwise very difficult and uncontrolled.

In this paper, we demonstrate that the collection of open string correlation functions conformally related to an open pair of pants in every 2D rational CFT (RCFT) yields an exact FP tensor with infinite bond dimensions. By tiling these correlators over a given manifold and summing over all intermediate states, including primaries and descendants, we obtain the 2D RCFT path integral. However, this tiling process leaves behind holes, which must be reconciled for the correlators to match with an FP tensor. Previous research \cite{Hung:2019bnq}   introduced shrinkable boundary conditions that address this problem and was further studied in \cite{Brehm:2021wev}. By combining these boundary conditions with the open correlators, we achieve a field theoretical construction of tensors that satisfy the expected properties of a FP tensor.
Practical numerical algorithm not only seeks an exact FP tensor, but it is a central problem to find also an efficient basis so that a minimal number of bond dimensions is needed to reproduce accurate results. 
 To show that our construction achieves both goals, we provide explicit numerical examples, focusing on the Ising, Yang Lee and tri-critical Ising models. Our results demonstrate convincingly that our proposed 
FP tensors can accurately recover the closed spectrum of the exact 2D CFT when tiling a cylinder. The boundary changing operator (BCO) achieves accuracies comparable to existing numerical FP tensors when truncated to similar bond dimensions. This demonstrates that the BCO basis serves as an optimal representation of the FP tensors. It validates the intuition that Virasoro descendants with higher conformal dimensions are naturally suppressed.

Moreover, our construction of FP tensors has an interpretation from the topological holographic principle \cite{Gaiotto:2020iye, Freed:2022qnc, Ji:2019jhk,Kong:2020cie,Kong:2020jne, Apruzzi:2021nmk, Chatterjee:2022kxb,  Freed:2018cec, Kong:2019byq, Bhardwaj:2017xup, Albert:2021vts, Vanhove:2018wlb, Aasen:2020jwb}. Specifically, we can re-express the path-integral constructed from the FP tensors as $Z_{CFT} = \langle \Omega | \Psi \rangle$ \cite{Chen:2022wvy}, where $|\Psi\rangle$ is the ground state wave-function of the Levin-Wen model \cite{Levin:2004mi}, or Turaev-Viro topological quantum field theory (TQFT) \cite{Turaev:1992hq}. i.e. The 2D CFT is expressed as a 3D path-integral with non-trivial boundary condition, the latter rendering the generalised symmetries (that includes usual group symmetries and other non-invertible symmetries) \cite{Gaiotto:2014kfa,Moore:1988qv, Verlinde:1988sn, Petkova:2000ip, Fuchs:2002cm, Frohlich:2006ch,Bhardwaj:2017xup, Chang:2018iay,Thorngren:2018bhj} of the CFT explicit.  

 \begin{figure}
    \centering
    \begin{subfigure}[b]{0.15\textwidth}
        \centering
        \includegraphics[scale=0.35]{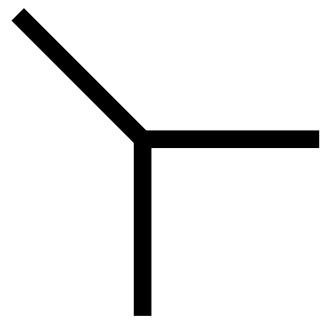}
        \caption{}
        \label{fig0a}
    \end{subfigure}
    \hfill
    \begin{subfigure}[b]{0.15\textwidth}
         \centering
        \includegraphics[scale=0.33]{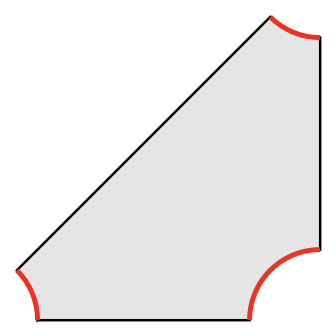}
        \caption{}
        \label{fig0b}
    \end{subfigure}
    \hfill
    \begin{subfigure}[b]{0.15\textwidth}
        \centering
        \includegraphics[scale=0.33]{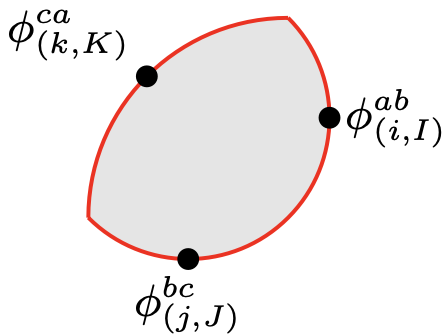}
        \caption{}
        \label{fig0c}
    \end{subfigure}
    \caption{(a) denotes the rank-3 tensor, corresponds to a path integral over the shaded region (b). (c) denotes correlation function of three local operators on a disk with conformal boundary condition on the red edge.  }
    \label{fig0}
\end{figure}

\section{Review of 2D conformal field theory}

CFT is a class of quantum field theories that are invariant under conformal transformations. These theories play a crucial role in statistical mechanics, condensed matter physics, and string theory. In 2D, CFTs are particularly rich due to the infinite-dimensional conformal symmetry algebra, making them highly solvable and powerful in understanding critical phenomena.

The conformal group in 2D is locally infinite-dimensional because transformations are governed by holomorphic and anti-holomorphic functions. The generators $L_n$'s of these conformal transformations satisfy the \textit{Virasoro algebra}, 
\begin{equation}
    [L_n,L_m]=(n-m)L_{n+m}+\frac{c}{12}(n^3-n)\delta_{n+m,0}.
\end{equation}
where $c$ is the so-caleld \textit{central charge}. 
\subsection{Primary operators, Virasoro descendants, and conformal dimension}

In 2D CFT, fields (or operators) depend on complex coordinates $z$ and $\bar{z}$. A primary field $O(z,\bar z)$ is characterized by two independent numbers $h$ and $\bar h$, which are the holomorphic and anti-holomorphic \textit{conformal dimension} (or scaling dimension). The total conformal dimension is given by:
\begin{equation}
    \Delta=h+\bar h. 
\end{equation}

The scaling behavior of a primary field under a dilation (scale transformation $z\rightarrow \lambda z$, $\bar z\rightarrow \lambda \bar z$) is:
\begin{equation}
    O(z,\bar z)\rightarrow \lambda^{\Delta} O(\lambda z,\lambda\bar z).
\end{equation}

Beyond primary fields, descendant fields arise from applying the Virasoro generators $L_{-n}$ and $\bar L_{-m}$. These descendants have higher conformal dimensions:
\begin{equation}
    h_{\text{descendant}}=h+n,\qquad \bar h_{\text{descendant}}=\bar h+m,
\end{equation}
where $n,m>0$ correspond to the levels of descendant operators.  

\subsection{OPE and conformal block}
The OPE states that when two operators $O_i(z,\bar z)$ and $O_j(w,\bar w)$ are close to each other, their product can be expanded as a sum over local operators. In the case of primary operators, we have
\begin{equation}
\begin{split}
    &O_i(z,\bar z)O_j(w,\bar w)\\
    =&\sum_k C_{ijk}(z-w)^{h_k-h_i-h_j}(\bar z -\bar w)^{\bar h_k-\bar h_i-\bar h_j}O_k(w,\bar w)\\
    &+\text{descendants}.
\end{split}
\end{equation}
where the coefficient $C_{ijk}$ is the \textit{structure constant}. It also appears in the three point function of local operators, 
\begin{equation}
    \langle O_i(z_1,\bar z_1)O_j(z_2,\bar z_2)O_k(z_3,\bar z_3) \rangle 
    =C_{ijk}\beta_{ijk}(z_{1,2,3})\bar\beta_{ijk}(\bar z_{1,2,3}),
\end{equation}
where $\beta_{ijk}$ is the three-point conformal block. For primary operators, it is,
\begin{equation}
\begin{split}
    &\beta_{ijk}(z_{1,2,3})=\\
    &\frac{1}{(z_1-z_2)^{h_1+h_2-h_3}(z_2-z_3)^{h_2+h_3-h_1}(z_1-z_3)^{h_1+h_3-h_2}}.
\end{split}
\end{equation}

\subsection{F-symbol and crossing symmetry}
In a CFT, the OPE determines how two operators interact. However, when dealing with three or more operators, we must specify the order in which we fuse them. Consider the fusion of three operators in two ways:
\begin{enumerate}
    \item First fuse $O_1$ and $O_2$ into $O_k$, then fuse $O_k$ with $O_3$:
    \begin{equation}
        (O_1O_2)O_3.
    \end{equation}
    \item Alternatively, first fuse $O_2$ and $O_3$ into $O_l$, then fuse $O_l$ with $O_1$:
    \begin{equation}
        O_1(O_2O_3).
    \end{equation}
\end{enumerate}
Since different fusion paths should give the same physical result, there must be a consistent way to transform between them. The coefficients governing this transformation are called the \textit{F-symbols} (or fusion matrix), denoted as $[F^{ijk}_l]^{blocks}_{mn}$. It is defined through the relation,
\begin{equation}\label{eq:Fsym}
    \sum_m[F^{ijk}_l]^{blocks}_{mn}C_{ijm}C_{klm}=C_{jkn}C_{iln}
\end{equation}

It encodes the associativity of the OPE, and also appears in the transformation between different fusion channels in a four-point function.

Consider the correlator:
\begin{equation}
    G(z,\bar z):=\langle O_i|O_j(1,1)O_k(z,\bar z)|O_l\rangle.
\end{equation}

It can be computed using the s-channel OPE:
\begin{equation}
    O_iO_j\rightarrow \sum_mC_{ijm}O_m, \qquad O_kO_l\rightarrow \sum_nC_{kln}O_n, 
\end{equation}
giving rise to the s-channel decomposition,
\begin{equation}
    G(z,\bar z)=\sum_m C_{ijm}C_{klm}\mathcal{F}^m_{ijkl}(z)\bar{\mathcal{F}}^m_{ijkl}(\bar z)
\end{equation}
where $\mathcal{F}^m(z)$ is the four-point conformal block. 

Alternatively, it can be computed using the t-channel OPE:
\begin{equation}
    O_jO_k\rightarrow \sum_pC_{jkp}O_p, \qquad O_iO_l\rightarrow \sum_qC_{ilq}O_q,
\end{equation}
leading to the t-channel decomposition, 
\begin{equation}
    G(z,\bar z)=\sum_p C_{jkp}C_{ilp}\mathcal{F}^p_{ijkl}(1-z)\bar{\mathcal{F}}_{ijkl}^p(1-\bar z).
\end{equation}

\textit{crossing symmetry} is the condition that a four-point function remains the same regardless of the order in which operators are fused. Equating the two decompositions, we have 
\begin{equation}
\begin{split}
    &\sum_m C_{ijm}C_{klm}\mathcal{F}_{ijkl}^m(z)\bar{\mathcal{F}}_{ijkl}^m(\bar z)\\
    =&\sum_p C_{jkp}C_{ilp}\mathcal{F}_{ijkl}^p(1-z)\bar{\mathcal{F}}_{ijkl}^p(1-\bar z).
\end{split}
\end{equation}

The change of bases from s-channel to t-channel four-point blocks is characterized by the F-symbol, 
\begin{equation}
    \mathcal{F}_{ijkl}^m(z)=\sum_n [F^{ijk}_l]^{blocks}_{mn} \mathcal{F}_{ijkl}^n(1-z).
\end{equation}

 This equation together with Eq.~\eqref{eq:Fsym} ensures the crossing symmetry relation. 

\subsection{Boundary conformal field theory}
Boundary CFT (BCFT) is a generalization of CFT defined on a manifold with a boundary, and the fields must satisfy certain boundary conditions that preserve a subset of the conformal symmetry.

In a bulk 2D CFT, the theory has the full conformal group. However, in the presence of a boundary, some conformal transformations are broken. In this setting, only conformal transformations that preserve the boundary survive. 

The presence of a boundary requires consistent boundary conditions on the fields. Boundary conditions must preserve the remaining conformal symmetry, leading to boundary states that satisfy the Virasoro constraints. For BCFT defined on the upper-half-plane (UHP), conformal invariance at the boundary (real axis) implies \cite{CARDY1984514}:
\begin{equation}\label{eq:gluing}
    T(z)=\bar T(\bar z) \quad \text{on the boundary } \Im(z)=0,
\end{equation}
which will be referred to as the \textit{gluing condition} for the energy momentum tensor. 

In standard CFT, local operators are inserted in the bulk. However, in BCFT, we also have boundary operators $\psi(x)$ inserted directly on the boundary. They are independent degrees of freedom living on the boundary.

The general boundary operators can interpolates between different boundary conditions. Consider such an operator inserted at $x$, where the boundary condition jumps from one type to another. These boundary operators are called \textit{boundary-changing operators} (BCO). They are denoted as $\psi_i^{ab}(x)$.  Before $x$, the system is in boundary condition $a$, while after $x$, the system is in boundary condition $b$. The BCO acts as a "sewing point" where the two boundary conditions meet \cite{CARDY1989581}.

Similar to the bulk CFT operators, BCOs satisfy boundary operator product expansions, 
\begin{equation}
    \psi_i^{ab}(x)\psi_j^{bc}(y)=\sum_k C_{ij}^{(abc)k}|x-y|^{h_k-h_i-h_j}\psi_k(y) +\text{descendants}. 
\end{equation}

The structure constant $C_{ij}^{(abc)k}$ now carries both operator indices $i,j,k$ and boundary condition indices $a,b,c$. They are related to the F-symbols which we discuss in detail in the \appref{app:Fsymbol}.  

\subsection{Boundary state formalism}

In each BCFT, one can associate a so-called \textit{boundary state} which incorporates the defining data of a boundary condition into objects built from Hilbert space of the original bulk CFT on the full complex plane. 

A boundary state, denoted as $\ket{a}\rangle$, corresponds to a specific  conformal boundary condition $a$, must satisfies the gluing condition in Eq.~\eqref{eq:gluing}. By applying an exponential mapping, this gluing condition translates into a constraint on the state, 
\begin{equation}
   \left( L_n-\bar L_{-n}\right) \ket{a}\rangle=0. 
\end{equation}

The solutions to this equation are linear combinations of a basis of states that are not normalisable with respect to the ordinary inner product, known as \textit{Ishibashi states} \cite{doi:10.1142/S0217732389000320, doi:10.1142/S0217732389000228}.   

These states are constructed from the primary fields of bulk CFT $\ket{\Delta}$, by acting with Virasoro generators, 
\begin{equation}
    \ket{\Delta}\rangle=\sum_{N=0}^{\infty}\sum_{j=1}^{d_h(N)}\ket{h,N;j}\otimes\overline{\ket{\bar h,N,j}}
\end{equation}
where $d_h(N)$ is the degeneracy of $N$-th level descendant states. 

While Ishibashi states satisfy the conformal invariance conditions at the boundary,  they do not correspond to local boundary conditions of the CFT path-integral. To construct a state corresponding to a local boundary condition, one must impose the Cardy’s condition. The condition essentially imposes that the trace over boundary-changing operators propagating in an annulus is equivalent to the overlap of two boundary states across a cylinder\cite{CARDY1989581}. i.e. 
\begin{equation} \label{eq:cardy}
\textrm{tr}_{ab} e^{-  \tau H_{ab}} = {}_c\langle a| e^{- H/\tau} | b\rangle_c ,
\end{equation}
where $H_{ab}$ is the Hamiltonian on a strip with left and right conformal boundary conditions $a$ and $b$ respectively. The states $ |a\rangle, |b\rangle$ are the corresponding boundary states of these boundary conditions, and $H$ is the Hamiltonian that describes the propagation of a closed circle along a cylinder.

 One could solve for the boundary states $|a\rangle$ in terms of the Ishibashi states defined above. 
This is a difficult problem in general. In the case where the CFT is a {\it diagonal rational CFT}, these boundary states could be solved explicitly, and they are known as \textit{Cardy's state}.
In a diagonal rational CFT there is a finite  number of families of primaries. Each primary operator has vanishing spin with both the chiral and anti-chiral component labeled by the same highest weight representation $i$ (of the chiral symmetry of the CFT, which includes Virasoro and other symmetries such as Kac-Moody etc. For minimal models the chiral symmetry includes only Virasoro symmetries.).
In this case, the label set of local conformal boundary conditions coincides with the set of primaries $\{i\}$.
The Cardy's boundary states are expressed as linear combinations of Ishibashi states as follows:
\begin{equation}
    \ket{i}_c=\sum_{j}\frac{S_{ij}}{\sqrt{S_{0j}}}\ket{j}\rangle.
\end{equation}

Here, the matrix $S_{ij}$, known as \textit{modular S-matrix}, encoding transformation properties of characters of the chiral symmetry under modular transformation \cite{CAPPELLI1987445, Verlinde:1988sn}. For further details on the S-matrix and BCFT, we refer the readers to \cite{Cardy:2008jc, DiFrancesco:1997nk}.

\section{The structure of FP Tensor}  
The FP tensor we propose, denoted as $\mathcal{T}^{abc}_{(i,I)(j,J)(k,K)}$, comprises nine indices. The labels $a$, $b$, $c$ correspond to the conformal boundary conditions of the RCFT, while $i$, $j$, $k$ represent the labels of the RCFT primaries, and the indices $I$, $J$, $K$ pertain to the descendants of their respective primaries.
In the RCFT, $a$, $b$, $c$ and $i$, $j$, $k$ take values from a finite set, while $I$, $J$, $K$ live in an infinite-dimensional space. Consequently, the exact FP tensors possess an infinite bond dimension, as expected. The FP tensor, $\mathcal{T}^{abc}_{(i,I)(j,J)(k,K)}$, can be interpreted as the path integral of the CFT within an open triangle. See \figref{fig0b}.
To regulate the path integral, we slightly modify the corners of the triangle and impose conformal boundary conditions labeled as $a$, $b$, and $c$ at each respective corner.  The edges of the triangle correspond to states that can be mapped to boundary-changing operators that connect the two conformal boundaries associated with the given edge (See \figref{fig0c}).  

 To show that they correspond to FP tensors, we need to demonstrate three properties:
(a) the FP tensors should satisfy crossing relations; (b) FP tensors covering a large patch upon contraction reproduce exactly the same FP tensors covering a smaller patch;
 (c) Tiling the FP tensors on a surface and assigning appropriate contraction of the indices recover the CFT path-integral on the surface.
These conditions (a) and (b) are illustrated in \figref{fig1} and \figref{fig2}. 
As we will see, these requirements ensure that the FP tensors reconstruct the CFT partition function exactly.

\begin{figure}[H]
    \centering
    \includegraphics[scale=0.37]{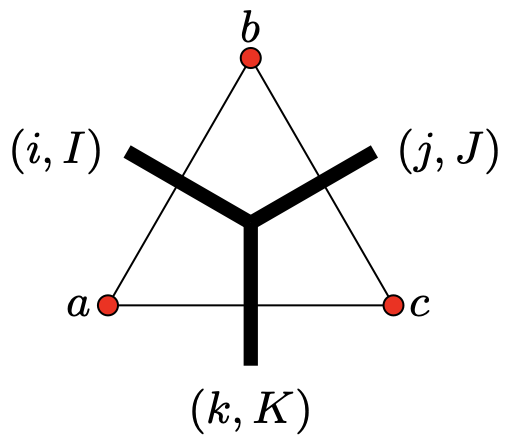}
    \caption{diagrammatical representation of the tensor. The base triangle denotes the structure coefficient $C^{abc}_{ijk}$, and the rank-3 tensor on top of it denotes the conformal block $\alpha^{ijk}_{IJK}$ which carries descendants information.   }
    \label{fig4}
\end{figure}

In general, the FP tensor can be decomposed as: 
\be \label{Tin6j}
\mathcal{T}^{abc}_{(i,I)(j,J)(k,K)} =  \alpha_{IJK}^{ijk}  C^{abc}_{ijk}
\ee
This is because a three-point correlation function of three boundary operators carries two parts, represented diagrammatically in \figref{fig4}, namely the structure coefficients $C^{abc}_{ijk}$ and the 3-point conformal blocks carrying the dependence of the correlation function on the precise descendant in the primary families, the location of insertions, and the precise shape of the manifold in which operators are inserted.
To set our notations, the three point correlation functions of three primary boundary-changing operators on the upper-half-plane (UHP) is given by:
 \begin{align}
&\langle O^{ab}_{(i,0)} (x_1) O^{bc}_{(j,0)}(x_2)O^{ca}_{(k,0)}(x_3) \rangle_{\text{UHP}} = C^{abc}_{ijk} \beta^{ijk}_{000} (x_1,x_2,x_3), \\
&\beta^{ijk}_{000} (x_1,x_2,x_3)= \\
& \frac{1}{|x_1- x_2|^{h_i + h_j - h_k}|x_1- x_3|^{h_i + h_k - h_j}|x_3- x_2|^{h_k + h_j - h_i}}, \nonumber
 \end{align}
where $I=J=K=0$ denotes the fact that the inserted operators are all primaries. Conformal blocks involving other descendants where $I,J,K\neq 0 $ can be generated by repeated use of the Virasoro or generally Kac-Moody operators in the primaries. 

\begin{figure}[H]
    \def\phione{\phi_i^{ab}}
    \def\phitwo{\phi_j^{bc}}
    \def\phithree{\phi_k^{ca}}
    \centering
    \begin{subfigure}[b]{0.2\textwidth}
        \centering
        \includegraphics[scale=0.40]{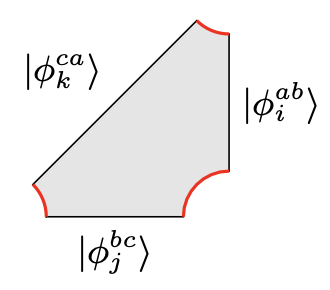}
        \caption{}\label{figA1a}
    \end{subfigure}
    \hfill
    \begin{subfigure}[b]{0.2\textwidth}
     \centering
     \includegraphics[scale=0.35]{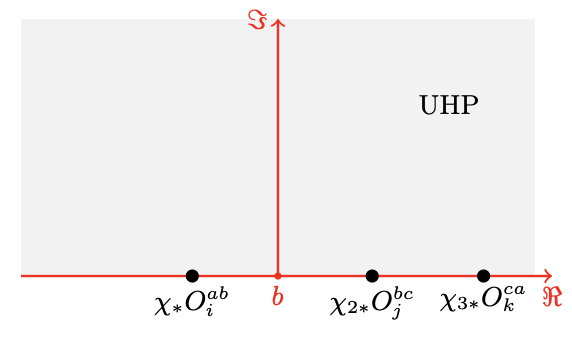}
    \caption{}\label{figA1b}
    \end{subfigure}
    \caption{Conformal transformation that maps the amplitude in triangular region to upper-half-plane with three local operators inserted on the real-axis. }
    \label{figA1}
\end{figure}

\begin{figure}[H]
    \centering
        \begin{subfigure}[b]{0.2\textwidth}
            \centering 
            \includegraphics[scale=0.40]{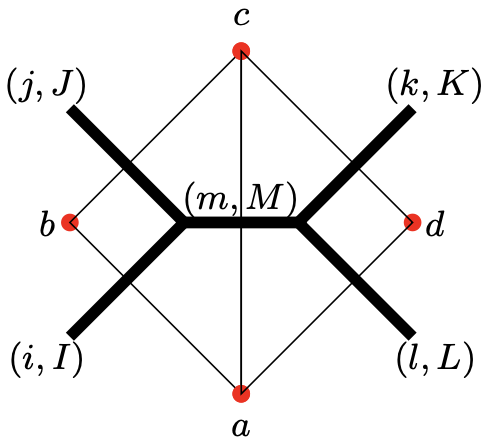}
        \end{subfigure}
        \hfill
        \begin{subfigure}[b]{0.2\textwidth}
            \centering
            \includegraphics[scale=0.40]{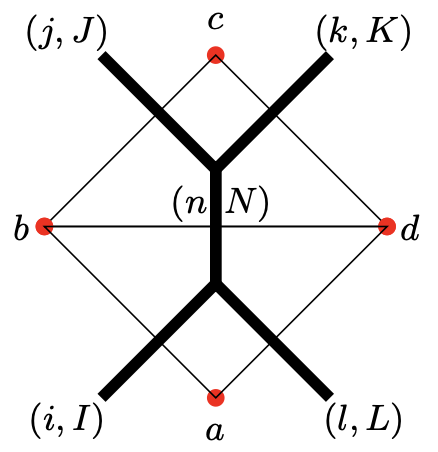}
        \end{subfigure}
    \caption{Crossing relation of FP tensor.}
    \label{fig1}
\end{figure}

In our proposed FP tensor, $\alpha_{IJK}^{ijk}$ is related to  $\beta_{IJK}^{ijk}$ by some conformal maps $\chi_{1,2,3}$, 
\begin{equation}
    C_{ijk}^{abc}\alpha_{IJK}^{ijk}=\langle \chi_{1*}O^{ab}_{(i,I)} (x_1) \chi_{2*}O^{bc}_{(j,J)}(x_2)\chi_{3*}O^{ca}_{(k,K)}(x_3) \rangle_{\text{UHP}}.
\end{equation}
where the conformal transformations $\chi_{1,2,3}$ map the amplitude on triangle to three point function on UHP, see \figref{figA1}. As we will explain with more detail in the next section, each $\chi_i$ is a composition of two conformal maps. The first maps the state on each edge in \figref{fig0b} to local operator in \figref{fig0c}, and the second maps the three point function on disk in \figref{fig0c} to UHP, allowing us to calculate it using $\beta_{IJK}^{ijk}$.  The coordinates $x_{1,2,3}=\chi_{1,2,3}(0)$ are fixed and suppressed in the following. 

These $\alpha^{ijk}_{IJK}$'s are called 3-point conformal blocks and fully determined by conformal symmetry. They satisfy:
\be \label{eq:crossing1}
\sum_{M}\alpha^{ijm}_{IJM}\alpha^{mkl}_{MKL} = \sum_{n,N}[F^{ijk}_l]^{\textrm{blocks}}_{mn}  \alpha^{inl}_{INL} \alpha^{jkn}_{JKN},
\ee
where $[F^{ijk}_l]^{\textrm{blocks}}$ are the crossing  coefficients characterising this RCFT. The same matrix $F^{\textrm{blocks}}$ also relate structure coefficients $C_{ijk}^{abc}$ through the equation \cite{CARDY1991274,LEWELLEN1992654},
\begin{equation}\label{eq:structure}
    \sum_m[F^{ijk}_l]^{\textrm{blocks}}_{mn} C_{ijm}^{abc}C_{mkl}^{acd}=C_{inl}^{abd}C_{jkn}^{bcd}.
\end{equation}
This guarantees the proposed FP tensor satisfies the crossing relation in condition (a).
\begin{align}
&\sum_{m,M}\mathcal{T}^{abc}_{(i,I)(j,J)(m,M)}\mathcal{T}^{acd}_{(m,M)(k,K)(l,L)}  \nonumber \\
&= \sum_{n,N} \mathcal{T}^{abd}_{(i,I)(n,N)(l,L)}\mathcal{T}^{bcd}_{(j,J)(k,K)(n,N)}.
\end{align}

\begin{figure}[H]
    \centering
    \includegraphics[scale=0.45]{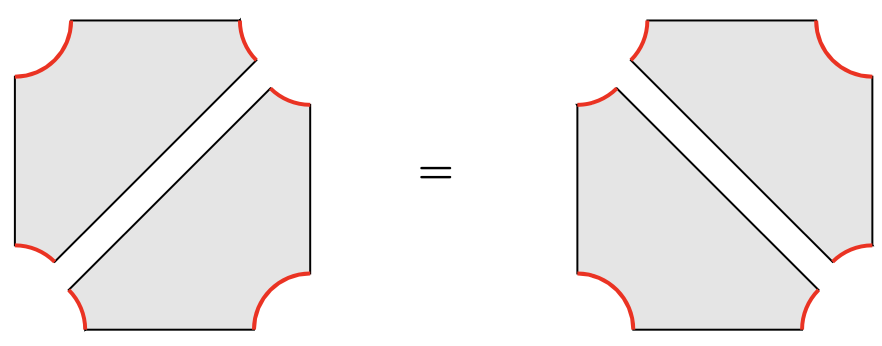}
    \caption{Crossing symmetry: two ways of gluing CFT path-integral on triangles are equivalent.}
    \label{fig1c}
\end{figure}

\begin{figure}[H]
    \centering
        \includegraphics[scale=0.45]{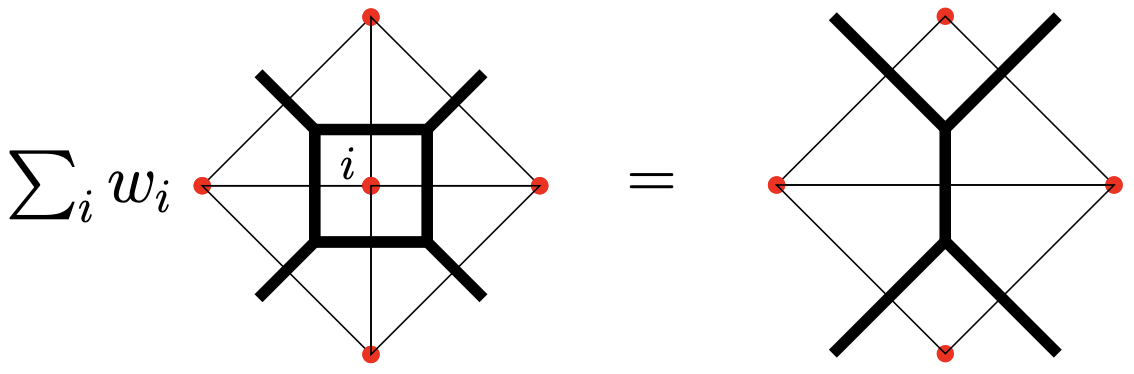}
    \caption{Coarse graining condition of FP tensor.}
    \label{fig2}
\end{figure}

Diagrammatically, this is illustrated in \figref{fig1}, which follows from the crossing symmetry of the RCFT (see \figref{fig1c}). 

The FP tensor also satisfies the coarse graining condition (b),
which is illustrated in \figref{fig2}. 
We note that the vertex degree of freedom at the center is summed over with a weight $w_{i}$. For a diagonal RCFT, 
\be \label{eq:weights}
w_{i} = S_{00}^{1/2} S_{i0},
\ee

\begin{figure}[H]
    \centering
    \includegraphics[scale=0.45]{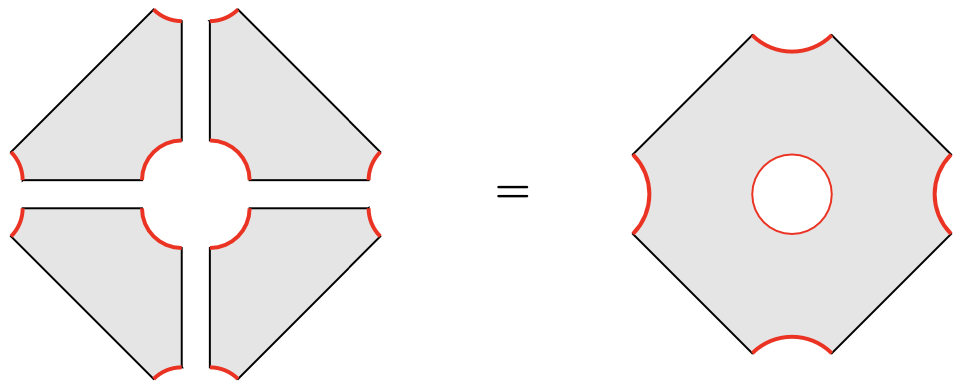}
    \caption{Gluing four triangles into a square with hole in the center.}
    \label{fig2c}
\end{figure}

Physically, the coarse-graining step corresponds to sewing four triangles by contracting the shared descendant labels between neighboring triangles along the shorter edges, as in \figref{fig2c}. However we note that a small hole is left in the middle. In order for the coarse-graining condition to be satisfied, this hole need to disappear. This can be done by first summing over the conformal boundary conditions with weights given by (\ref{eq:weights}), followed by shrinking the size of hole to zero. The idea of this weighted sum of conformal boundary conditions was initially explored in \cite{Hung:2019bnq}. It was called the entanglement brane boundary conditions, but perhaps more appropriately, the shrinkable boundary condition, where these boundaries were designed to shrink and disappear, reproducing a smooth path-integral.  Another motivation is that the weighted sum makes the boundary transparent to the topological defect in CFT \cite{Brehm:2021wev}.  These considerations motivated the use of this particular weighted sum.

The open boundary can be transformed through a modular transformation into a closed conformal boundary Cardy state $|i\rangle_c$. It can be shown that the weighted sum of the boundaries yields:
\be \label{eq:closing}
\sum_i w_i |i\rangle_c = |0\rangle\rangle,
\ee
where the right-hand side corresponds to the identity of the Ishibashi state. When the size of hole $r$ is small, the state is evolved by a long Euclidean time and becomes, 
\begin{equation}
\begin{split}
    &\lim_{r\rightarrow 0} e^{-H|\log r|}|0\rangle\rangle\\
    =&e^{\frac{c}{6}\pi|\log r|}\left(|0\rangle+\frac{2}{c}e^{-8\pi|\log r|}L_{-2}\bar L_{-2}|0\rangle+\cdots\right).
\end{split}
\end{equation}

The dominant contribution arises from the leading term, which is the vacuum state. The leading corrections then come from the leading descendant of the vacuum state, which can be viewed as an irrelevant perturbation in the thermodynamic limit of the tiling, as explained in \cite{Brehm:2021wev}. This boundary conditions are physical reasons behind condition (b) and (c) satisfied by the FP tensor. 

The partition function of the CFT on a manifold $M$ can be obtained using the following procedure. We begin by triangulating the manifold $\mathcal{M}$ into a collection of triangles ${\triangle}$. Each edge $e$ on a triangle is labeled with a pair of primary and descendant labels $(i,I)$, and each vertex $v$ is labeled with a conformal boundary condition $a$. On each triangle, we assign a tensor $\mathcal{T}^{abc}_{(i,I)(j,J)(k,K)}$ based on the labeling of the edges and vertices. The proposed partition function is then given by:
\begin{align}
Z_{M} =  \sum_{\{ (i,I)\}, \{a\} } \prod_v \omega_{a}    
\prod_{\triangle}\mathcal{T}^{a b c}_{(i,I) (j,J)(k,K)}.  \label{eq:partition}
\end{align}

\section{Computing the Conformal block} 
\begin{figure}[H]
    \def\phione{\phi_i^{ab}}
    \def\phitwo{\phi_j^{bc}}
    \def\phithree{\phi_k^{ca}}
     \centering
     \begin{subfigure}[b]{0.25\textwidth}
        \includegraphics[scale=0.45]{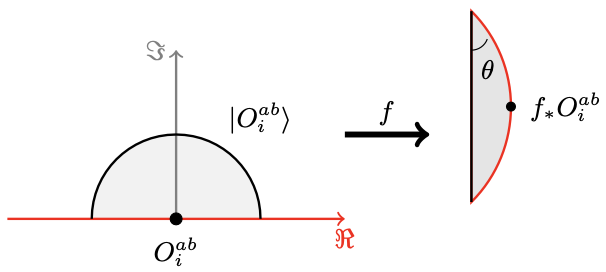}
         \caption{}
         \label{figA2b}
     \end{subfigure}
      \begin{subfigure}[b]{0.2\textwidth}
        \centering
        \includegraphics[scale=0.45]{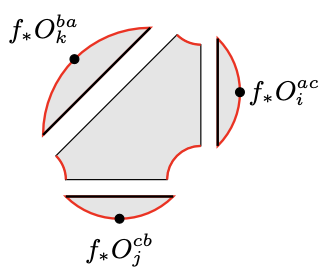}
        \caption{}
        \label{figA2a}
    \end{subfigure}
      \caption{(a)  The semi-disk is mapped to a circular segment region through a function $f(z)$. (b) The amplitude of three open states can be calculated by attaching three segment regions along the open boundaries of triangle. }
     \label{figA}
 \end{figure}

To calculate each component of the FP tensor, we evaluate the conformal block $\alpha_{IJK}^{ijk}$, which is calculated by the path-integral on the triangular region.  In this article, we provide two methods, one is trough the state-operator correspondence.  Another method can be found at \appref{app:pants}. The standard state-operator correspondence allows us to prepare a state $|\opi\rangle$ on the unit semi-circle by inserting local operator $\opi$ at the origin. See the left side of \figref{figA2b}. Now we find a function $f(z)$ which maps the semi-disk to a circular segment region, as shown in  right side of \figref{figA2b}. The operator $\opi$ is mapped to $f_*\opi$ inserted on the arc of the segment.  Through this map we prepare some state $|\phi_i^{ab}\rangle$ on the vertical edge of the segment region.

To calculate the amplitude in \figref{figA1a} we attach these prepared states to the three open boundaries of the triangular region. Diagrammatically, this process is represented by gluing the segments along the three open boundaries as shown in \figref{figA2a}. 

\begin{figure}[H]
    \centering
    \includegraphics[scale=0.35]{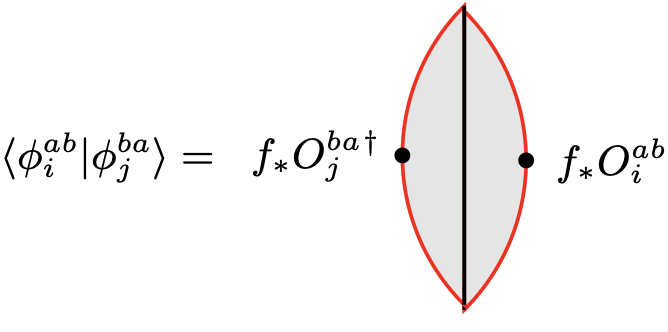}
    \caption{The double segment region obtained by gluing two circular segments along the vertical edge.}
    \label{figA3}
\end{figure}

We further require that the prepared states form an orthonormal basis in the Hilbert space of boundary CFT. The inner product of these states is determined by the two-point function on the nut-shaped region as shown in \figref{figA3}. 

\begin{figure}[H]
    \centering
    \includegraphics[scale=0.35]{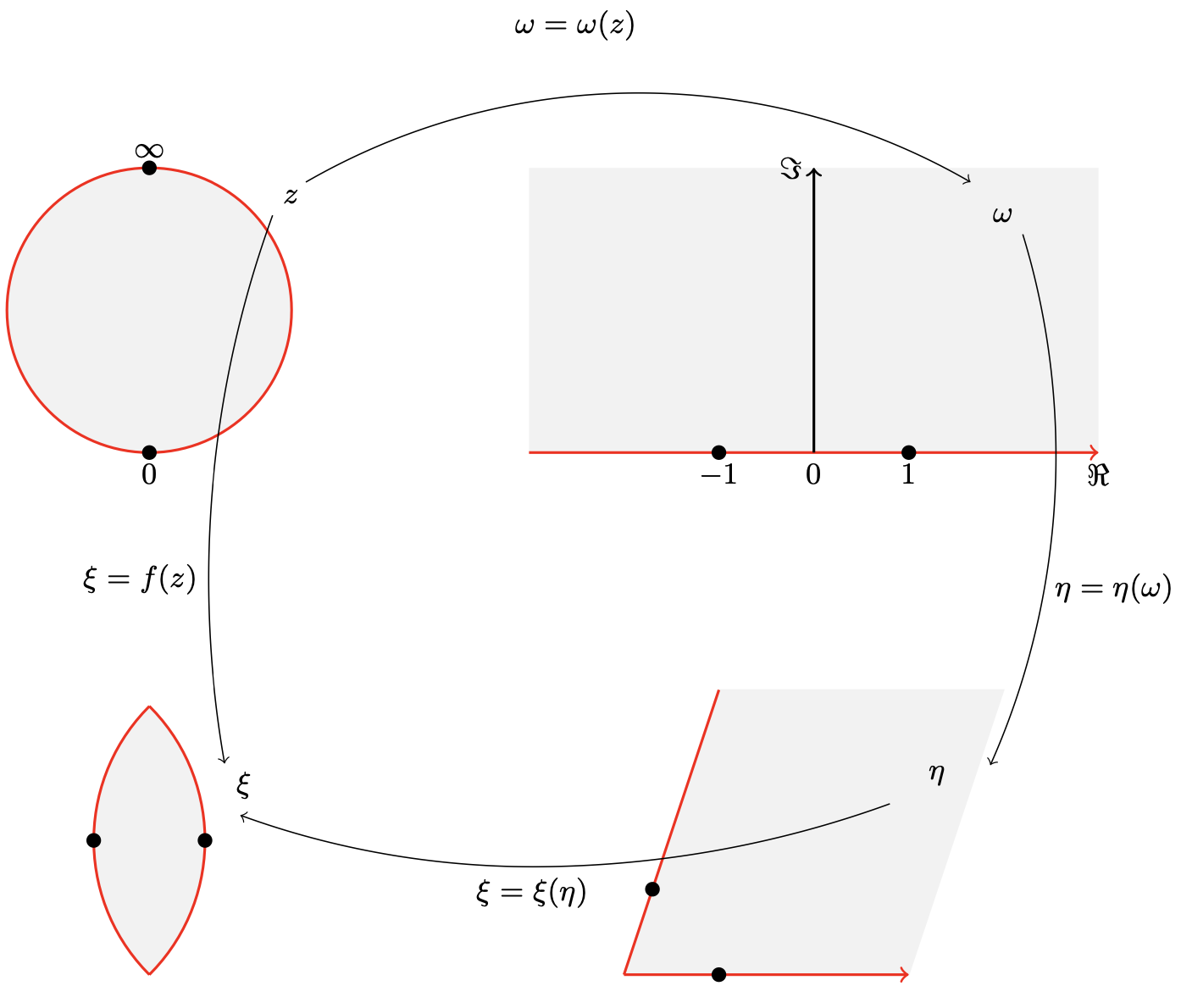}
    \caption{Conformal map from compactified UHP (top left) to the nut region (bottom left) as composition of three simple maps. Black dots represent location of the operators insertion.}
    \label{Afig5}
\end{figure}

The condition of orthonormality is expressed through the following equation of two point function on the nut: 
\begin{equation}\label{eqotho}
\langle [f_{*}O_j^{\dagger}](-b)[f_{*}O_i](b)\rangle_{\text{nut}}=\delta_{ij}.
\end{equation}
where $b$ is the coordinates of operators on the nut. 
The two-point function can be readily calculated by observing that the same function $f(z)$ maps the compactified UHP to the nut-shaped region. This mapping is a composition of three simple transformations, whose effects are illustrated in  \figref{Afig5},
\begin{equation}
\begin{split}
   & f(z)=\xi\circ\eta\circ\omega(z)\\
  &  \omega(z)=\frac{1+z}{1-z},\ \  \eta(\omega)=e^{-i\theta}\omega^{\frac{2\theta}{\pi}},\ \ \xi(\eta)=i\frac{\eta-1}{\eta+1}.
\end{split}
\end{equation}

Since $f(z)$ mapps UHP to the nut, the two point function of operator $f_*\opi$ on the nut is equal to the two point function of operator $\opi$ on the UHP. 
\begin{equation}\label{eqotho}
\langle [f_{*}{\op{j}{b}{a}}^{\dagger}](-b)[f_{*}\opi](b)\rangle_{\text{nut}}=\langle{\op{j}{b}{a}}^{\dagger}(\infty)\opi(0)\rangle_{\text{UHP}}.
\end{equation}

Therefore the condition in Eq.~\eqref{eqotho} is equivalent to finding orthonormal set of basis operators $\opi$. 

Here we have a free parameter $\theta$ which is the angle of the corner in the segment region (see \figref{figA2b}). This angle serves as a gauge freedom of our tensor construction. For computation simplicity we choose to set $\theta=\frac{\pi}{4}$. 

\begin{figure}[H]
     \def\phione{\phi_i^{ab}}
    \def\phitwo{\phi_j^{bc}}
    \def\phithree{\phi_k^{ca}}
    \centering
    \begin{subfigure}[b]{0.2\textwidth}
        \centering
        \includegraphics[scale=0.55]{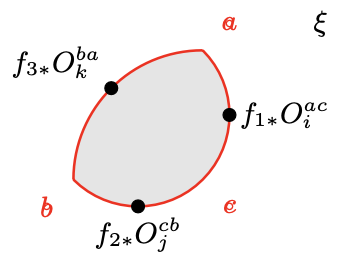}
        \caption{}\label{figA5a}
    \end{subfigure}
    \hfill
    \begin{subfigure}[b]{0.2\textwidth}
     \centering
     \includegraphics[scale=0.45]{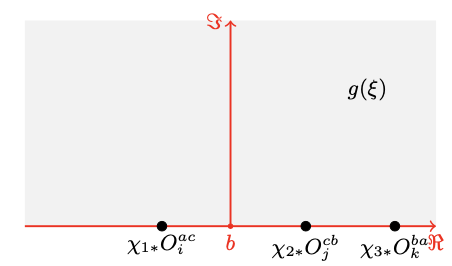}
    \caption{}\label{figA5b}
    \end{subfigure}
    \caption{Conformal transformation $g(\xi)$ maps the disk with two cusps (a) to the upper-half-plane (b). }
    \label{figA5}
\end{figure}

Following the state to operator map, we proceed by shrinking the corners of the triangular region to zero length. This leads to the disk with smooth boundary except for the two cusps.  Three operators are inserted on the boundary as shown in \figref{figA5a}. 

 Note that although the red arcs at the corners of the triangle shrink to zero size, they still carry conformal boundary conditions. These conditions determine both the structure constant of the triangle amplitude and the set of allowed BCOs inserted on the disk boundary. When the triangles are glued together to form partition function, these boundary conditions must be summed with the correct weights, as prescribed in Eq.~\eqref{eq:closing}.


From the previous discussion, we see that the orthonormal basis states on the triangle edges are prepared by the transformed BCOs, $f_*O^{ab}_i$, inserted at the boundary of the nuts. To insert these operators on the disk boundary, as shown in \figref{figA5a}, we employ three functions,  $f_1$, $f_2$ and $f_3$, which are related to $f$ by translation and rotation according to the orientation of the state in \figref{figA2a}. 
\begin{equation}
\begin{split}
   & f_1(z)=f(z)+1, \ \ f_2(z)=-if(z)-i\\
   & f_3(z)=\sqrt{2}e^{i\frac{3\pi}{4}}f(z)
\end{split}
\end{equation}
They map the origin of \figref{figA2b} to the three points on the disk in \figref{figA5a}. 

 In the final step we find a function $g(\xi)$ that maps the disk (\figref{figA5a}) to the UHP (\figref{figA5b}). Given the gauge choice $\theta=\frac{\pi}{4}$, the disk boundary has only two cusps. We choose the function $g(\xi)$ which maps them to $0$ and $\infty$. This function be written down explicitly as,
\begin{equation}
    g(\xi)=(-i\frac{\xi+(1+i)}{\xi-(1+i)})^{\frac{4}{3}}.
\end{equation}

 After applying the conformal transformation $g(\xi)$, the operators on the UHP in \figref{figA5b} becomes  $\chi_{1*}\op{i}{a}{c}(x_1)$, $\chi_{2*}\op{j}{c}{b}(x_2)$ and $\chi_{3*}\op{k}{b}{a}(x_3)$, 
 where the $\chi$-functions are defined as the following composition of the mapping $f$ and $g$, 
\begin{equation}\label{eqhs}
    \begin{split}
        &\c_1(z)=g(f(z)+1),\ \ \c_2(z)=g(-if(z)-i), \\
        &\c_3(z)=g(\sqrt{2}e^{i\frac{3\pi}{4}}f(z)).
    \end{split}
\end{equation}

They map the origin of \figref{figA2b} to three points on the real axis in \figref{figA5b}, with coordinates $x_1=\chi_1(0)$, $x_2=\chi_2(0)$, $x_3=\chi_3(0)$. So the tensor components are equal to the following three point functions on upper-half-plane,

\begin{equation}
\scalebox{0.85}{
    $\mathcal{T}^{acb}_{(I,i)(J,j)(K,k)}=\langle \c_{1*}O_{(I,i)}^{ac}(x_1) \c_{2*}O_{(J,j)}^{cb} (x_2) \c_{3*}O_{(K,k)}^{ba}(x_3)\rangle_{UHP}$,
    }
\end{equation}

As a simple example, consider the three boundary operators all being primary fields with  conformal dimension $h_1$, $h_2$ and $h_3$. The dependence on conformal boundary condition $a$,$b$,$c$ are only contained in the structure constant $C^{abc}_{ijk}$. We  express the conformal block $\alpha_{IJK}^{ijk}$ as, 
\begin{equation}\label{eq:primarytensor}
\begin{split}
&\alpha^{h_1h_2h_3}_{000} \\
=&\frac{|\chi_1'(0)|^{h_1}|\chi_2'(0)|^{h_2}|\chi'_3(0)|^{h_3}}{|x_1-x_2|^{h_1+h_2-h_3}|x_2-x_3|^{h_2+h_3-h_1}|x_1-x_3|^{h_1+h_3-h_2}}\\
\approx& 0.266^{h_1+h_2}0.704^{h_3}\\
=&0.515^{\Delta_1+\Delta_2}0.839^{\Delta_3},
\end{split}
\end{equation}
where $\Delta_i=2h_i$ are the corresponding bulk conformal dimensions.

\section{Examples}
\subsection{The Ising CFT}
In the Ising example, we can put in explicit expressions to the above construction. The closed Ising CFT has three primaries $\mathcal{C}_{Is} =\{I, \psi,\sigma\}$. The theory has three conformal boundary conditions. They are labeled as $\{+,-, f\}$, corresponding to the respective primaries. The Hilbert space for an interval with left and right boundary given by $a$ and $b$ respectively, where $a,b \in \mathcal{C}_{Is}$ is given by
$\mathcal{H}_{ab} = \oplus_c  N_{ab}^c V_c$,  
where $V_c$ is the space corresponding to the primary representation labeled $a\in \mathcal{C}_{Is}$, and $N_{ab}^c \in \mathbb{Z}_{\ge 0}$ are the fusion coefficient among the objects  $\mathcal{C}_{Is}$, with: 
$
N_{I b}^c = \delta_{bc}, \,\,\, N_{\sigma \sigma}^{c} = 1 - \delta_{c\sigma}, \,\,\, N_{\sigma a\neq \sigma}^{b} = \delta_{b\sigma}, \,\,\, N_{\psi \psi}^b = \delta_{bI}.
$

The matrix $[F^{ijk}_l]^{\textrm{blocks}}$ is provided in \appref{app:Fsymbol}.  Using Eq.~\eqref{eq:structure}, the structure coefficients $C^{abc}_{ijk}$ can be calculated \cite{LEWELLEN1992654}.  Below we list those values other than $1$:
\begin{align}
   &\T{\pm\pm\pm}{I I I}=\T{\pm\mp\pm}{\pp \pp I}=\T{\pm f\pm}{\si\si I}=2^{\frac{1}{4}}, \\
    & \T{\pm f \mp}{\si\si\pp}=\frac{1}{2^{\frac{1}{4}}}, \T{f+f}{\si\si\pp}=\frac{1}{\sqrt{2}}, \T{f-f}{\si\si\pp}=-\frac{1}{\sqrt{2}} \nonumber
\end{align}

\begin{figure}
   \centering
    \includegraphics[scale=0.65]{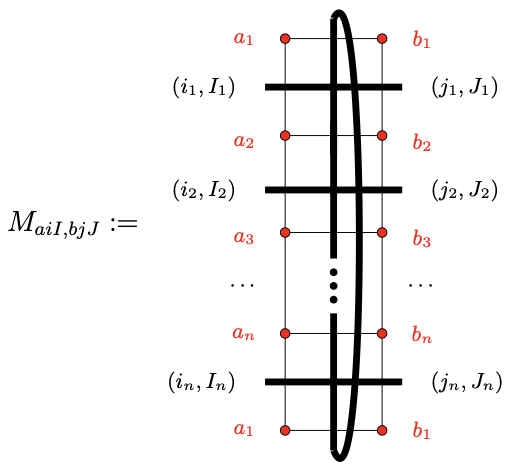}
       \caption{The transfer matrix}
       \label{figA10}
\end{figure}

We have to compute the three point functions involving descendants, and then transform them into the needed geometry using the conformal map constructed in the last section.
Explicitly, one has to first look for the orthogonal basis of the descendants. 
For example, in level one, the normalized first descendant $O^{(-1)}$ is defined as $\frac{1}{\sqrt{2h}}L_{-1}O$. It's transformation under the conformal map $\chi(z)$ is:
\begin{equation}
    \chi_*[O^{(-1)}]=|\chi'(0)|^{h}\left(\chi'(0)O^{(-1)}+\sqrt{\frac{h}{2}}\frac{\chi''(0)}{\chi'(0)}O\right).
\end{equation}
In the second level, we find three normalized operators,
\begin{align}
& \I^{(-2)}=2L_{-2}\I,\\
& \psi^{(-2)}= \frac 6 {25} L_{-2}\psi+\frac 9 {25} L_{-1}^2\psi, \\
& \si^{(-2)}=\frac{16\sqrt{2}}{25} L_{-2}\si+\frac{12\sqrt{2}}{25} L_{-1}^2\si.
\end{align}
and the corresponding transformation rules given by: 
\begin{align}
&\chi_*[L_{-2}O] = (\chi')^2  L_{-2}O + \frac{3}{2} \chi'' L_{-1}O + \nonumber \\
&\left(\frac{c \chi'''}{12\chi'} - \frac{c (\chi'')^2}{8 (\chi')^2} + \frac{2\chi''' h }{3\chi'} - \frac{(\chi'')^2 h }{4(\chi')^2}\right) O, \\
\nonumber
&\chi_* [L_{-1}^2O] =(\chi')^2L_{-1}^2O+(2h+1)\chi''L_{-1}O+\\
&\left(\frac{\chi'''h}{\chi'}+\frac{(\chi'')^2h(h-1)}{(\chi')^2}\right)O.
 \end{align}
For higher level descendants, we derive recursive equations to solve all the transformation rules. Additionally, the three-point correlation functions for descendant fields are also calculable by recursive methods.  The details are provided in the \appref{app:recursive}. 

With all the ingredients we numerically calculate the tensors and check the crossing relations (a) and coarse-graining condition (b), truncating each conformal family to a finite number of descendants, up to a maximum conformal dimension of $h_{max}=5$.  The numerical results are presented in Table \ref{tab:crossing}, and Table \ref{tab:components} in the \appref{app:numerical}.  Despite the small bond dimension, we find that these conditions hold with an accuracy of $2 \times 10^{-3}$.

Finally, We demonstrate that our proposed FP tensor constructed from open correlation functions can indeed recover the closed spectrum with
surprisingly high accuracy despite keeping only very few descendants in each family. The cylinder is constructed using 4 squares formed out of 8 triangles, as shown in \figref{figA10}. The labels of the conformal boundaries at the left and the right edge of the cylinders are treated alongside the primaries and descendant labels of the FP tensors as input and output indices of the cylinder. In our notation, the transfer matrix is denoted by $M_{aiI,bjJ}$. The indices $aiI$ is collective representation of all the conformal boundary labels, primary labels and descendant labels. That is $a=\{a_1,a_2,\cdots, a_n\}$, $i=\{i_1,i_2,\cdots,i_n\}$ and  $I=\{I_1,I_2,\cdots,I_n\}$.
One can solve for the spectrum of the cylinder, which is listed in the \tabref{tab1}. In the \appref{app:transfer} we provide details on how the numerical conformal dimension converges towards the precise value as we increase the cutoff in descendant level.

\begin{table}
\centering
\begin{tabular}{||c|c|c||}
    \hline
    States & Numerical dim & Accurate dim \\
    \hline 
    $\I$ & 0.0000 & 0.0000\\
    $\si$ & 0.1250 & 0.1250\\
    $\psi$ & 0.9989 & 1.0000\\
    $\partial\si, \bar{\partial}\si$ & 1.1253 & 1.1250 \\
    $\partial\psi, \bar{\partial}\psi$  & 2.0004 & 2.0000 \\
    $\I^{(-2)}$, $\bar{\I}^{(-2)}$ & 1.9986 & 2.0000\\
    $\partial\bar{\partial}\sigma$ & 2.1099 & 2.1250\\
    $\si^{(-2)}$, $\bar{\si}^{(-2)}$ & 2.1208 & 2.1250\\
    $\partial\bar{\partial}\psi$ & 2.9517 & 3.0000\\
    $\partial^2\psi$, $\bar{\partial}^2\psi$ & 3.0030 & 3.0000\\
    $\I^{(-3)}, \bar{\I}^{(-3)}$ & 2.9496 & 3.0000 \\
    $\bar{\partial}\si^{(-2)},\partial \bar{\si}^{(-2)}$ & 2.9900 & 3.1250\\
    $\partial \si^{(-2)}$, $\bar{\partial}\bar{\si}^{(-2)}$ & 3.1167 & 3.1250 \\
    $\sigma^{(-3)},\bar{\sigma}^{(-3)}$ & 3.1291 & 3.1250\\
    $\I^{(-2,-\bar{2})}$ & 4.0323 & 4.0000\\
    \hline
\end{tabular}
\caption{Ising model: conformal dimensions from fixed-point tensor vs. exact value. Cylinder length L=4, descendant level cut=5.  }
\label{tab1}
\end{table}

The central charge can also be obtained, as explained in \appref{app:transfer}. Here we list the central charge value vs. the descendant level cut-off in Table \ref{tab:central}.
\begin{table}[]
    \centering
    \begin{tabular}{||c|c|c|c|c|c|c||}
    \hline
      descendant & 0 & 1 & 2 & 3 & 4 & exact\\
      \hline
        $c_{\text{Ising}}$ &0.4565 &0.4900& 0.5048  & 0.5033 & 0.4975 & 0.5\\
        \hline
    \end{tabular}
    \caption{Ising model: central charge computed from FP tensor as function of the descendant level cut-off. }
    \label{tab:central}
\end{table}

\subsection{The Yang-Lee CFT and tri-critical Ising CFT}
In this section, we present numerical results for Yang-Lee CFT and tri-critical Ising CFT. Note that Yang-Lee is a non-unitary CFT, with negative state norm and negative conformal dimension. This renders the some of the tensor components to take complex value.  Despite this we still able to recover the bulk state conformal spectrum by diagonalizing the transfer matrix. 

Yang-Lee model has two primary operators, denoted by $\I$ and $\tau$, with conformal dimension $0$ and $-\frac{2}{5}$. The fusion rule is $\tau\times\tau=\I+\tau$. We also label the corresponding conformal boundary condition by $\{\I,\tau\}$. The structure constants are given below, 
\begin{equation}
\begin{split}
   & C_{III}^{III}=1,\ C_{III}^{\tau\tau\tau}=-1.272i,  \ C_{\tau\tau I}^{\tau\tau\tau}=-1.272i \\ 
   &C_{\tau\tau I}^{\tau I \tau}=-1.272i, \ C_{\tau\tau I}^{I\tau I}=1, \ C_{\tau\tau\tau}^{\tau I\tau}=-1.560, \\ &C_{\tau\tau\tau}^{\tau\tau\tau}=-2.523.
\end{split}
\end{equation}

In Table \ref{tab:YLspectrum} we list the bulk conformal dimension obtained from the FP tensor and compare them with the exact values.  

\begin{table}
\centering
\begin{tabular}{||c|c|c||}
\hline
States & \text{Numerical dim} & \text{Accurate dim} \\ 
\hline
$\tau$ &-0.403 & -0.400 \\ 
$\I$&0.000 & 0.000 \\ 
$\partial\tau$, $\bar{\partial}\tau$ & 0.598 & 0.600 \\ 
$\tau^{(2)}$,$\bar{\tau}^{(2)}$& 1.591 & 1.600 \\ 
$\partial\bar{\partial}\tau$ & 1.600 & 1.600 \\ 
$\I^{(2)}$, $\bar{\I}^{(2)}$& 1.996 & 2.000 \\ 
$\bar{\partial}\tau^{(2)}$, $\partial\bar{\tau}^{(2)}$ &2.530 & 2.600 \\ 
$\tau^{(3)}$, $\bar{\tau}^{(3)}$ & 2.594 & 2.600 \\ 
$\I^{(3)}$, $\bar{\I}^{(3)}$ & 2.939 & 3.000 \\ 
\hline
\end{tabular}
\caption{Yang-Lee model: conformal dimension from fixed point tensor vs. exact value. Cylinder length L=6, descendant cut at level 3. }
\label{tab:YLspectrum}
\end{table}

For tri-critical ising CFT, the primary fields are labeled by $\I$, $\Phi$, $\Psi$, $\Xi$, $\Lambda$ and $\Omega$, with corresponding conformal dimension $\{0,\frac{1}{10}, \frac{3}{5}, \frac{3}{2}, \frac{7}{16}, \frac{3}{80}\}$. The structure constants can be calculated using Eq.~\eqref{eq:boundaryope} and Eq.~\eqref{eq:bulkope} and listed in Table \ref{tab:TCIS}. We calculate the eigenvalues of the transfer matrix, listed in Table \ref{tab:spectrum_TCIS}, and compare them with the bulk conformal dimension. 

\begin{table}
\centering
\scalebox{0.8}{
\begin{tabular}{||c|c|c|c|c|c|c|c|c|c|c|c|c|c|c|c|c|c|c|c||}
\hline
$C_{\Phi\Phi\I}^{\Phi\I\Phi}$ & 0.7862 & $C_{\Phi\Phi\I}^{\Phi\Psi\Phi}$ & 0.7862 & $C_{\Phi\Phi\Psi}^{\Phi\Psi\Phi}$ & -0.4281 & $C_{\Phi\Phi\I}^{\Psi\Phi\Psi}$ & 0.7862 & $C_{\Phi\Phi\I}^{\Psi\Xi\Psi}$ & 0.7862\\  $C_{\Phi\Phi\Psi}^{\Psi\Phi\Psi}$ & 0.4281 & $C_{\Phi\Phi\Psi}^{\Psi\Xi\Psi}$ & -0.6927 & $C_{\Phi\Psi\Phi}^{\Psi\Xi\Phi}$ & 0.6927 & $C_{\Phi\Psi\Xi}^{\Psi\Xi\Phi}$ & 0.5147 & 
$C_{\Psi\Psi\I}^{\Psi\I\Psi}$ & 0.7862 \\ 
$C_{\Psi\Psi\I}^{\Psi\Psi\Psi}$ & 0.7862 & $C_{\Psi\Psi\Psi}^{\Psi\Psi\Psi}$ & -0.4281 & $C_{\Phi\Phi\I}^{\Omega\Lambda\Omega}$ & 0.6611 & $C_{\Phi\Phi\I}^{\Omega\Omega\Omega}$ & 0.6611 & $C_{\Phi\Phi\Psi}^{\Omega\Lambda\Omega}$ & -0.5825 \\ $C_{\Phi\Phi\Psi}^{\Omega\Omega\Omega}$ & 0.36 & $C_{\Phi\Psi\Phi}^{\Omega\Lambda\Omega}$ & 0.5825 & $C_{\Phi\Psi\Xi}^{\Omega\Omega\Omega}$ & 0.4328 &
$C_{\Phi\Psi\Xi}^{\Omega\Lambda\Omega}$ & -0.4328 &$C_{\Psi\Psi\I}^{\Omega\Omega\Omega}$ & 0.6611 \\ $C_{\Psi\Psi\I}^{\Omega\Lambda\Omega}$ & 0.6611 & $C_{\Psi\Psi\Psi}^{\Omega\Omega\Omega}$ & -0.36 & $C_{\Psi\Psi\Psi}^{\Omega\Lambda\Omega}$ & 0.5825 & $C_{\Phi\Omega\Lambda}^{\Omega\Lambda\Phi}$ & -0.4674 & $C_{\Phi\Omega\Lambda}^{\Omega\Omega\Phi}$ & 0.4674 \\ $C_{\Phi\Omega\Omega}^{\Omega\Omega\Phi}$ & 0.4409 & $C_{\Phi\Omega\Lambda}^{\Omega\Omega\Psi}$ & 0.4674 & 
$C_{\Phi\Omega\Lambda}^{\Omega\Lambda\Psi}$ & 0.4674 & $C_{\Phi\Omega\Omega}^{\Omega\Omega\Psi}$ & -0.4409 & $C_{\Psi\Omega\Omega}^{\Omega\Omega\Psi}$ & -0.18 \\ $C_{\Psi\Omega\Omega}^{\Omega\Lambda\Psi}$ & -0.2912 & $C_{\Psi\Omega\Lambda}^{\Omega\Lambda\Psi}$ & 0.5725 & $C_{\Lambda\Lambda\I}^{\Lambda\I\Lambda}$ & 0.8409 & $C_{\Lambda\Lambda\I}^{\Lambda\Xi\Lambda}$ & 0.8409 & $C_{\Lambda\Lambda\Xi}^{\Lambda\Xi\Lambda}$ & -0.8409 \\ $C_{\Lambda\Lambda\I}^{\Omega\Phi\Omega}$ & 0.6611 & 
$C_{\Lambda\Lambda\I}^{\Omega\Psi\Omega}$ & 0.6611 & $C_{\Lambda\Lambda\Xi}^{\Omega\Psi\Omega}$ & -0.6611 & $C_{\Lambda\Lambda\Xi}^{\Omega\Phi\Omega}$ & 0.6611 & $C_{\Omega\Omega\I}^{\Omega\I\Omega}$ & 0.6611 \\ $C_{\Omega\Omega\I}^{\Omega\Psi\Omega}$ & 0.6611 & $C_{\Omega\Omega\I}^{\Omega\Phi\Omega}$ & 0.6611 & $C_{\Omega\Omega\I}^{\Omega\Xi\Omega}$ & 0.6611 & $C_{\Omega\Omega\Psi}^{\Omega\Xi\Omega}$ & 0.2912 & $C_{\Omega\Omega\Phi}^{\Omega\Xi\Omega}$ & -0.7134 \\ 
$C_{\Omega\Omega\Xi}^{\Omega\Xi\Omega}$ & -0.0883 & $C_{\I\I\I}^{\I\I\I}$ & 1.0 & $C_{\I\I\I}^{\Phi\Phi\Phi}$ & 0.7862 & $C_{\I\I\I}^{\Psi\Psi\Psi}$ & 0.7862 & $C_{\I\I\I}^{\Xi\Xi\Xi}$ & 1.0 \\ $C_{\I\Phi\Phi}^{\Xi\Xi\Psi}$ & 1.0 & $C_{\I\Psi\Psi}^{\Xi\Xi\Phi}$ & 1.0 & $C_{\I\Xi\Xi}^{\Xi\Xi\I}$ & 1.0 & $C_{\I\I\I}^{\Lambda\Lambda\Lambda}$ & 0.8409 & 
$C_{\I\I\I}^{\Omega\Omega\Omega}$ & 0.6611 \\ $C_{\I\Phi\Phi}^{\Lambda\Lambda\Omega}$ & 0.8409 & $C_{\I\Psi\Psi}^{\Lambda\Lambda\Omega}$ & 0.8409 & $C_{\I\Xi\Xi}^{\Lambda\Lambda\Lambda}$ & 0.8409 & $C_{\I\Xi\Xi}^{\Omega\Omega\Omega}$ & 0.6611 & $C_{\Phi\Psi\Xi}^{\Lambda\Omega\Lambda}$ & 0.5505\\ 
\hline
\end{tabular}
}
\caption{Structure Constant of tri-critical Ising model.}\label{tab:TCIS}
\end{table}

\begin{table}
\centering
\begin{tabular}{||c|c|c||}
\hline
States & Numerical dim & Accurate dim \\ 
\hline
$\I$ &0.0000 & 0.000 \\ 
$\Omega$& 0.075  & 0.075 \\ 
$\Phi$&0.200  & 0.200 \\ 
$\Lambda$&0.873  & 0.875 \\ 
$\partial \Omega$, $\bar{\partial} \Omega$& 1.076  & 1.075 \\ 
$\Psi$ & 1.196  & 1.200 \\ 
$\partial \Phi$, $\bar\partial \Phi$& 1.201  & 1.200 \\ 
$\partial\Lambda$, $\bar\partial\Lambda$&1.871  & 1.875 \\ 
$\I^{(2)}$, $\bar\I^{(2)}$&2.013  & 2.000 \\ 
$\partial\bar\partial\Omega$& 2.064  & 2.075 \\ 
$\Omega^{(2)}$, $\bar{\Omega}^{(2)}$&2.084  & 2.075 \\ 
$\Omega^{(2)'}$, $\bar{\Omega}^{(2)'}$&2.087  & 2.075 \\ 
$\partial\bar\partial \Phi$& 2.190  & 2.200 \\ 
$\Phi^{(2)}$, $\bar\Phi^{(2)}$&2.192  & 2.200 \\ 
$\partial\Psi$, $\bar\partial\Psi$& 2.210  & 2.200 \\ 
$\partial\bar\partial \Lambda$ &2.877  & 2.875 \\ 
$\Lambda^{(2)}$, $\bar\Lambda^{(2)}$&2.878  & 2.875 \\
$I^{(3)}$, $\bar I^{(3)}$& 2.978 & 3.000 \\
$\Xi$& 3.004  & 3.000 \\ 
\hline
\end{tabular}
\caption{Tri-critical Ising model: conformal dim from fixed point tensor vs. exact value. Cylinder length L=3, descendant cut at level 2.  }\label{tab:spectrum_TCIS}
\end{table}

Finally we list the value of central charge computed from the transfer matrix in Table \ref{tab:charge} for both of the two models.

\begin{table}[]
    \centering
    \begin{tabular}{||c|c|c|c|c|c|c||}
    \hline
      descendant   & 0 & 1 & 2 & 3 & 4 & exact \\
      \hline
       $c_{\text{TCIS}}$ & 0.4024 & 0.6540 & 0.6820 & 0.6964 & 0.6996 &0.7  \\
       \hline
       $c_{\text{YL}}$ & -4.7307& -4.4143 & -4.4029 & -4.3997 & -4.3998 & -4.4\\
       \hline
    \end{tabular}
    \caption{Central charge computed from FP tensor for Yang-Lee (YL) model and tri-critical Ising (TCIS) model }
    \label{tab:charge}
\end{table}

\section{FP tensors as eigenstates of topological RG operators}
While the FP tensor can be understood directly as a CFT correlation function without explicit reference to an associated 3d TQFT, 
it is an important observation that these FP tensors follows from an exact eigenstate of the topological RG operator \cite{Chen:2022wvy,Vanhove:2018wlb}, and the CFT partition function can be written explicitly as a {\it strange correlator}. 

To appreciate this connection, recall that the label set of primaries in an RCFT are objects in a modular fusion category $\mathcal{C}$. Here we focus on diagonal RCFT so that the conformal boundary conditions are also labeled by objects in $\mathcal{C}$.  It is convenient to re-scale the three point conformal block $\alpha^{ijk}_{IJK} =\mathcal{N}_{ijk}\,\gamma^{ijk}_{IJK}$, where \cite{Kojita:2016jwe} ,
\be\label{eq:Ns}
\mathcal{N}_{ijk} = \sqrt{\theta(i,j,k)/\sqrt{d_id_jd_k}},
\ee
 where $\theta(i,j,k) = d_i/\left[F^{jkkj}\right]^{\textrm{blocks}}_{1i}$, and   $d_i$ is the quantum dimension of object $i$, which is related to the modular matrix by $d_i = S_{0i}/S_{00}$ for a diagonal RCFT. The value of the FP tensor (\ref{Tin6j}) does not change, except that it is decomposed instead as  $\mathcal{T}^{abc}_{(i,I)(j,J)(k,K)} = \gamma^{ijk}_{IJK} \hat C^{abc}_{ijk}$.  On this basis, the structure coefficients $\hat C^{abc}_{ijk}$ of a diagonal RCFT (including the Ising CFT described above) can be written simply as \cite{Fuchs:2004xi}, 
\be\label{eq:structure}
\hat C^{abc}_{ijk} = (d_id_jd_k)^{1/4}  \left[ \begin{tabular}{ccc}
$ i$ &$j$ & $k$\\
  $c$&$a$&$b$\end{tabular}\right],
\ee
where the square bracket denotes the quantum 6j-symbols of the modular tensor category $\mathcal{C}$ associated to the RCFT in with tetrahedral symmetry and chosen normalization. Several components in this gauge involving the identity label are fixed to the values reviewed in the \appref{app:Fsymbol}. 
All two point correlations are also normalised. 
These $\gamma^{ijk}_{IJK}$ inherit the crossing relation of (\ref{eq:crossing1}), with the crossing kernel re-scaled as: 
\be
[F^{ijk}_l]_{mn} = [F^{ijk}_l]^{\textrm{blocks}}_{mn} \frac{\mathcal{N}_{jkn}\mathcal{N}_{inl}}{\mathcal{N}_{ijm}\mathcal{N}_{mkl}}.
\ee
These re-scaled crossing kernals $ [F^{ijk}_l]_{mn}$ is related to the quantum 6j-symbol above by:
 \be
 [F^{ijk}_l]_{mn} = \sqrt{d_m d_n} \left[ \begin{tabular}{ccc}
$ i$ &$j$ & $m$\\
  $k$&$l$&$n$
 \end{tabular}\right ].
 \ee
 The explicit values of $ [F^{ijk}_l]_{mn}$ and $ [F^{ijk}_l]^{\textrm{block}}_{mn}$ for the Ising CFT are given in the \appref{app:Fsymbol}. Now it should be obvious that (\ref{eq:partition}) can be rewritten as a strange correlator 
$
Z_M= \langle \Omega | \Psi\rangle,
$
where $ | \Psi\rangle$ is the ground state of the Levin-Wen model corresponding to the fusion category $\mathcal{C}$.  It is well known that such a wave-function on a two dimension surface 
can be constructed using the Turaev Viro formulation of TQFT path-integral over a triangulated three ball \cite{Turaev:1992hq}. 
For a surface triangulation that matches the tiling as specified in (\ref{eq:partition}), the Levin-Wen ground state wavefunction can be written as \cite{Turaev:1992hq,wavefunction1,wavefunction2}:
\begin{align}
|\Psi\rangle = \sum_{\{a_v\}} \sum_{\{i\}}  \prod_{e}  d_{i}^{1/2} \prod_v \omega_{a} \prod_{\triangle} \begin{bmatrix}
    i & j & k  \\
    c & a & b
    \end{bmatrix}  |\{i\}  \rangle, 
\end{align}

\begin{figure}
    \begin{subfigure}{0.23\textwidth}
    \centering
    \includegraphics[scale=0.55]{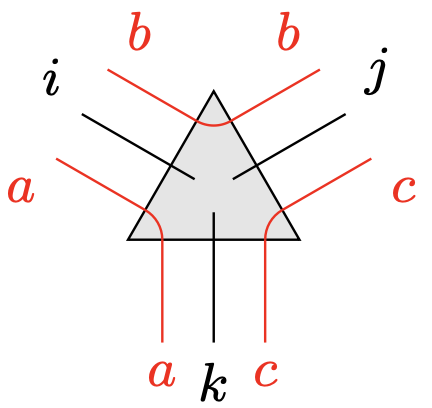}
    \caption{}
    \end{subfigure}
    \hfill
    \begin{subfigure}{0.23\textwidth}
    \centering
    \includegraphics[scale=0.60]{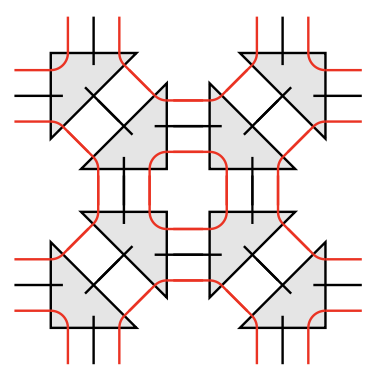}
     \caption{}
    \end{subfigure}
    \caption{(a) Triple-line tensor that describes the Levin-Wen ground state. (b) Tiling the tensors into a tensor network to represent the wavefunction $|\Psi\rangle$.}
    \label{fig:stringnet}
\end{figure}
The ket $|\{i\}\rangle$ are basis states 
living on the edges which carries a label $i \in \mathcal{C}$, and
\be
\langle \Omega| =  \sum_{\{(i, I)\}} \langle \{i\} |   \prod_{\triangle} \gamma_{IJK}^{ijk} .
\ee
The crossing relation (\ref{eq:crossing1}), together with (\ref{eq:closing}) guarantees that $\langle \Omega| $ is an eigenstate of the RG operator proposed in \cite{Chen:2022wvy}. 
We note that the entanglement brane boundary condition (\ref{eq:closing}) follows simply from the prescription of the Turaev-Viro formation of the path-integral. The weights assigned to each internal edge that is summed agrees with the weighted sum of the Cardy states in (\ref{eq:closing}). In other words, the associated 3d TQFT constructed from $\mathcal{C}$ knows about how to close holes in the RCFT. 

When constructing non-diagonal RCFTs, 
the boundary conditions of the CFT correspond to corner variables placed on triangles, which are generally labeled by objects from a "module category" $\mathcal{M}_{\mathcal{C}}$ associated with the fusion category $\mathcal{C}$. According to the TQFT framework \cite{Turaev:1992hq}, the corner variable should be summed with the weights given by the quantum dimension of the label as an object in the module category. This summation procedure yields the appropriate shrinkable boundary conditions for general RCFTs.
The strange correlator representation of the exact two-dimensional CFT partition function serves as an explicit, practical, and easily computable realization of the holographic relationship between a quantum field theory with categorical symmetry and a TQFT in one higher dimension, as advocated in Ref. \cite{Gaiotto:2020iye, Freed:2022qnc, Ji:2019jhk,*Kong:2020cie,*Kong:2020jne, *Chatterjee:2022kxb,  Apruzzi:2021nmk, Freed:2018cec, Kong:2019byq, Bhardwaj:2017xup, Albert:2021vts, Vanhove:2018wlb, Aasen:2020jwb}.

\section{Conclusion and discussion} 
In conclusion, we identify the field theoretic explanation of FP tensors that describes RCFT. Specifically, the FP tensor is shown to be a precise correlation function of RCFT involving "boundary-changing operators" defined on triangles. Our proposed construction of the FP tensor naturally fulfills all the requirements of real space RG conditions.  Importantly, not only do we argue that these correlation functions are exact FP tensors in the infinite bond dimension limit, they are also in an {\it optimal} basis. i.e. Contribution of descendants decays rapidly, so that keeping a small number of descendants produce a {\it finite dimensional} efficient approximation of the FP tensor.  The reproduction of the
lower bulk states spectrum and the satisfaction of all RG conditions to accuracy comparable to existing tensor network renormalisation algorithms at similar bond dimension serves as a robust validation of our approach.
This approach thus provides not merely a theoretical explanation of the FP tensors, but a practical way of simulating any (rational) CFT efficiently even if they did not follow from well known lattice models that offer a natural tensor network representation.  
We further back up the claim that the constructed FP tensor is in an optimal basis by explicitly showing that the tensor components match well with those obtained from tensor network renormalization method. This is illustrated in \appref{app:gauge} by making appropriate gauge transformation.


 Finally, one very important question is whether the fixed point tensor in higher dimensions also admit interpretation field theoretically. As emphasized in the paper, the exact fixed point tensor constructed in the current paper is related to the ``topological holographic principle", in which a three dimensional bulk emerges from the BCFT correlation functions. The emergent bulk essentially makes the generalised symmetry explicit. We note that the topological holographic principle applies quite generally to arbitrary dimensions \cite{Gaiotto:2020iye, Freed:2022qnc, Ji:2019jhk,*Kong:2020cie,Kong:2020jne, Chatterjee:2022kxb,  Apruzzi:2021nmk, Freed:2018cec, Kong:2019byq, Bhardwaj:2017xup, Albert:2021vts}. It has also been shown in explicit examples that well known lattice models, such as the 2+1 D Ising spin model, can readily be expressed as $Z_{3D\, Ising} = \langle \Omega| \Psi_{4D\, TQFT}\rangle$, and that the tensor network renormalization procedure in 3D can again be deconstructed in a way that leaves the 4D topological bulk explicit at every step, exactly as in the case for 2D lattice models by keeping the generalized symmetry explicitly preserved at every RG step\cite{Chen:2022wvy}.  This strongly suggests that the construction advocated in this paper should have a generalization in higher dimensions. The basic building blocks of 4D TQFT requires input of a 2-fusion category, and that would suggest that the field theoretic interpretation of the FP tensors would involve both codimension 1 and codimension 2 defects. This is in contrast to the 2D CFT situation where only codimension 1 defects (i.e. conformal boundaries) are involved. Therefore in some sense, we  expect that FP tensors in higher dimensions correspond to boundary-changing correlation functions in string field theory. 
Complete understanding of 2-categories, and the computation and classification of codimension 2 defects in field theories are important problems. That the FP tensors can be computed numerically offers exciting possibilities of amalgamating these results and to building up a topological bootstrap framework, which could potentially lead us towards a re-formulation of field theories algebraically, perhaps in all dimensions.

 \section{Acknowledgments} We acknowledge useful discussions with Xiaogang Wen, Shuheng Shao, Yikun Jiang, Bingxin Lao, Nicolai Reshetikhin, Gabriel Wong and Xiangdong Zeng.  This work is supported by funding from Hong Kong’s Research Grants Council (GRF no.14301219, no.14303722 and RGC Research Fellow Scheme 2023/24, No. RFS2324-4S02) 
 LYH acknowledges the support of NSFC (Grant No. 11922502, 11875111). LC acknowledges the support of NSFC (Grant No. 12305080) and the start up funding of South China University of Technology. GC acknowledges the support from Commonwealth Cyber Initiative at Virginia Tech, U.S. Department of Energy, Office of Science, Office of Advanced Scientific Computing Research. 

\appendix

\section{Convention for 6j-symbols and F symbols}\label{app:Fsymbol}

The crossing kernals $[F^{ijk}_l]_{mn}$ after being rescaled in the main text, which are often also referred to as the Racah coefficients in the literature, are related to quantum 6j-symbols as follows:
\begin{equation}
    F_{m n}\left[ \begin{footnotesize}\begin{tabular}{ccc}
$ j$ & $k$ \\
  $i$ & $l$\end{tabular}  \end{footnotesize} \right ]= \sqrt{d_md_n}\left[\begin{tabular}{ccc}
$ i$ & $j$ & $m$ \\
  $k$ & $l$ & $n$\end{tabular}\right].
\end{equation}
The quantum 6j symbols denoted by object in square brackets, enjoy full tetrahedral symmetry. 
In this gauge it fixes a number of components to:
\begin{equation}
    \left[ \begin{tabular}{ccc}
$ a$ &$a$ & $0$\\
  $b$&$b$&$c$\end{tabular}\right ]=
  \left[ \begin{tabular}{ccc}
$ a$ &$b$ & $c$\\
  $b$&$a$& $0$\end{tabular}\right ]=\frac{N^c_{ab}}{\sqrt{d_ad_b}}.
\end{equation}
Correspondingly, 
\begin{equation}
    [F^{aab}_{b}]_{0c}=\sqrt{\frac{d_c}{d_ad_b}}.
\end{equation}


As described in the main text, the Racah coefficients are related to the crossing kernels describing crossing relations between canonically normalised conformal blocks by a re-scaling. 

For the Ising CFT, the $F^{\text{blocks}}$ are given by the expressions below with the parameter $\lambda =1/2$:
\begin{eqnarray}
F_{11}\left[ \begin{footnotesize}\begin{tabular}{ccc}
$ \psi$ & $\psi$ \\
  $\psi$ & $\psi$\end{tabular}  \end{footnotesize} \right ] &=&1,\\
F_{11}\left[ \begin{footnotesize}\begin{tabular}{ccc}
$ \sigma$ & $\sigma$ \\
  $\sigma$ & $\sigma$\end{tabular}  \end{footnotesize} \right ] &=&
-F_{\psi\psi}\left[ \begin{footnotesize}\begin{tabular}{ccc}
$ \sigma$ & $\sigma$ \\
  $\sigma$ & $\sigma$\end{tabular}  \end{footnotesize} \right ] =\frac{1}{\sqrt{2}},\\
F_{1\psi}\left[ \begin{footnotesize}\begin{tabular}{ccc}
$ \sigma$ & $\sigma$ \\
  $\sigma$ & $\sigma$\end{tabular}  \end{footnotesize} \right ]&=& \frac{\lambda}{\sqrt{2}},\;\;\;
F_{\psi 1}\left[ \begin{footnotesize}\begin{tabular}{ccc}
$ \sigma$ & $\sigma$ \\
  $\sigma$ & $\sigma$\end{tabular}  \end{footnotesize} \right ]= \frac{1}{\sqrt{2}\lambda},\\
 F_{1\sigma}\left[ \begin{footnotesize}\begin{tabular}{ccc}
$ \psi$ & $\sigma$ \\
  $\psi$ & $\sigma$\end{tabular}  \end{footnotesize} \right ]&=&
  F_{1\sigma}\left[ \begin{footnotesize}\begin{tabular}{ccc}
$ \sigma$ & $\psi$ \\
  $\sigma$ & $\psi$\end{tabular}  \end{footnotesize} \right ]=\lambda, \\
F_{\sigma 1}\left[ \begin{footnotesize}\begin{tabular}{ccc}
$ \psi$ & $\psi$ \\
  $\sigma$ & $\sigma$\end{tabular}  \end{footnotesize} \right ]&=&
  F_{\sigma 1}\left[ \begin{footnotesize}\begin{tabular}{ccc}
$ \sigma$ & $\sigma$ \\
  $\psi$ & $\psi$\end{tabular}  \end{footnotesize} \right ]=\frac{1}{\lambda},\\
F_{\sigma \sigma}\left[ \begin{footnotesize}\begin{tabular}{ccc}
$ \psi$ & $\sigma$ \\
  $\sigma$ & $\psi$\end{tabular}  \end{footnotesize} \right ]&=&
  F_{\sigma \sigma}\left[ \begin{footnotesize}\begin{tabular}{ccc}
$ \sigma$ & $\psi$ \\
  $\psi$ & $\sigma$\end{tabular}  \end{footnotesize} \right ]=-1.
\end{eqnarray}

The Racah  coefficients of the Ising model 
are given by the same expressions above with $\lambda=1$.  The corresponding  6j symbols are given by
\begin{align}
\left[\begin{tabular}{ccc}
$ \sigma$ & $\sigma$ & $1$ \\
  $\sigma$ & $\sigma$ & $1$\end{tabular}\right]&=
\left[\begin{tabular}{ccc}
$ \sigma$ & $\sigma$ & $1$ \\
  $\sigma$ & $\sigma$ & $\psi$\end{tabular}\right]=\frac{1}{\sqrt{2}},
\left[\begin{tabular}{ccc}
$ \sigma$ & $\sigma$ & $\psi$ \\
  $\sigma$ & $\sigma$ & $\psi$\end{tabular}\right]=\frac{-1}{\sqrt{2}},\\
\left[\begin{tabular}{ccc}
$ 1$ & $1$ & $1$ \\
  $\sigma$ & $\sigma$ & $\sigma$\end{tabular}\right]&=
  \left[\begin{tabular}{ccc}
$ 1$ & $\psi$ & $\psi$ \\
  $\sigma$ & $\sigma$ & $\sigma$\end{tabular}\right]=2^{-\frac{1}{4}}.
\end{align}
One can readily check that the they are indeed related to $F^{\textrm{blocks}}$ by a rescaling of the form 
\be
[F^{ijk}_l]_{mn} = [F^{ijk}_l]^{\textrm{blocks}}_{mn} \frac{\mathcal{N}_{jkn}\mathcal{N}_{inl}}{\mathcal{N}_{ijm}\mathcal{N}_{mkl}}.
\ee

 In the main text,  the structure constants  $C_{ijk}^{abc}$ we listed, are related to $\hat C_{ijk}^{abc}$ through the equation, 
\begin{equation}\label{eq:boundaryope}
    C_{ijk}^{abc}=\frac{\hat C_{ijk}^{abc}}{\mathcal{N}_{ijk}}.
\end{equation}

To obtain these structure constant, we first use Eq.~\eqref{eq:structure} to calculate $\hat C_{ijk}^{abc}$, where we utilize the formula in \cite{Kirillov:1991ec} to compute the quantum 6j-symbol.  The factor $\mathcal{N}_{ijk}$'s, defined in Eq.~\eqref{eq:Ns}, are related to bulk structure constant trough the following equation \cite{Kojita:2016jwe}, 
\begin{equation}\label{eq:bulkope} \mathcal{N}_{ijk}=1/\sqrt{C^{\text{bulk}}_{ijk}}.
\end{equation}

For minimal models, the bulk structure constant can be computed  using the general formula in \cite{DOTSENKO1985291}.


\section{Another choice of conformal map -- pants-diagram}\label{app:pants}

\begin{figure}[H]
    \centering
    \includegraphics[scale=0.45]{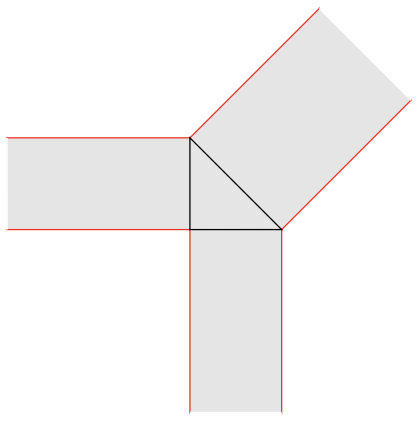}
    \caption{pants-diagram}
    \label{figA9}
\end{figure}

 In this section we give another conformal map that helps to calculate tensor as three point functions on the UHP. 
 In this approach we prepare the states from asymptotic infinity and evolve them in Euclidean time to the triangle boundary. Thereby we extend the triangular region as a pants-diagram, as shown in \figref{figA9}. The function $\chi$ is constructed by a map from pants-diagram to UHP.  

First we adopt the Schwarz–Christoffel transformation to find the map from upper-half-plane to the pants-diagram,  
\begin{equation}
    s(\xi)=\int^{\xi} dx\frac{\sqrt{2}(5x^2-1)^{\frac 1 4}}{x(x^2-1)}.
\end{equation}

This function maps the three points $-1$, $0$ and $1$ of the UHP to three infinities along the legs of pants, where we attach free open string states.  Near these infinities, we have the following expansions:
\begin{equation}
\begin{split}
    &s|_{\xi\rightarrow 1}\sim \ln |\xi-1|, \ \ s|_{\xi\rightarrow 0}\sim -(1+i)\ln|\xi|+i\pi, \\
    &s|_{\xi\rightarrow -1}\sim i\ln|\xi+1|+\pi.
\end{split}
\end{equation}

These are precisely the functions we can utilize to prepare open string states at infinities. According to these relations, we define the conformal functions, 
\begin{equation}
\begin{split}
    &\chi_1(z):=s^{-1}\left(\ln z\right), \ \chi_2(z):=s^{-1}\left(i\ln z+\pi\right),\\
    &\chi_3(z):=s^{-1}\left(-(1+i)\ln z+i\pi\right). 
\end{split}
\end{equation}

It's hard to find a concise expression for the inverse function of the map $s(\xi)$. Instead, we expand this function around the singularties. This gives us the series expansion for $\chi_i$'s around $z=0$ as, 
\begin{equation}
    \begin{split}
        &\chi_1(z)=1+0.9z+0.709z^2+0.641z^3+O(z^4)\\
        &\chi_2(z)=-1+0.9z-0.709z^2+0.641z^3+O(z^4)\\
        &\chi_3(z)=0.410z+0.008z^3+O(z^5)
    \end{split}
\end{equation}
This allows us to evaluate the tensor numerically. Again, we present the tensor component for  primary fields following the steps in \eqref{eq:primarytensor}
\begin{equation}
    \alpha^{\Delta_1\Delta_2\Delta_3}_{000} \approx 0.671^{\Delta_1+\Delta_2}0.905^{\Delta_3}.
\end{equation}

This tensor also reproduces CFT bulk spectrum. Therefore It is related to the construction in the main text by a gauge transformation. 

\section{Numerical verification for fixed-point condition}\label{app:numerical}

To test the fixed point property of proposed FP tensor, we calculated the tensor values explicitly for descendant fields up to conformal dimension $h_{max}=5$.  

For the crossing symmetry condition (see \figref{fig1}), we provide an example where we fix the four external legs as $(\s,\s,\s,\s)$ and the four boundary conditions as $(f,-,f,+)$. The following contractions are computed, 
\begin{equation}
\begin{split}
    &T_{L.H.S.}:= T^{f-f}_{\s\s\I}T^{f+f}_{\s\s\I}+T^{f-f}_{\s\s\psi}T^{f+f}_{\s\s\psi}\\
    &T_{R.H.S.}:= T^{+f-}_{\s\s\psi}T^{-f+}_{\s\s\psi},
\end{split}    
\end{equation}
where we didn't write the descendant field indices and they are understood as being contracted implicitly. For example,

\begin{equation}
    \left(T^{f-f}_{\s\s\I}T^{f+f}_{\s\s\I}\right)_{IJKL}:=\sum_M T^{f-f}_{(\s,I)(\s,J)(\I,M)}T^{f+f}_{(\s,K)(\s,L)(\I,M)}.
\end{equation}

In Table \ref{tab:crossing}, we list some tensor components of $T_{L.H.S.}$ vs. $T_{R.H.S.}$.

For the coarse-graining condition (see \figref{fig2}), consider the example of fixing the four external legs as $(\I,\I,\I,\I)$ and the four boundary conditions to be $(+,+,+,+)$, we compute the following contraction of four tensors as the the left-hand-side of the equation in \figref{fig2}: 
\begin{equation}
   T_{L.H.S.}:= [(T^{+++}_{\I\I\I})^4+(T^{+-+}_{\psi\psi\I})^4+\sqrt{2}(T^{+f+}_{\si\si\I})^4]/2\sqrt{2}. 
\end{equation}

Again we didn't write the descendant field indices and they are understood as being contracted implicitly according to \figref{fig2}. Similarly the right-hand-side of this equation is obtained by contracting two tensors,
\begin{equation}
\begin{split}
    T_{R.H.S.}=&(T^{+++}_{\I\I\I})^2\\
    =&\sum_M T^{+++}_{(\I,I)(\I,J)(\I,M)}T^{+++}_{(\I,K)(\I,L)(\I,M)}.
\end{split}
\end{equation}

In the Table \ref{tab:components} we present the numerical value of some tensor components. Despite the very small bond dimension, we find that they are satisfied to an accuracy of $2 \times 10^{-3}$. 

\begin{table}[]
    \centering
    \begin{tabular}{||c|c|c||}
\hline
Components & $T_{L.H.S.}$ & $T_{R.H.S.}$ \\ 
\hline
0000 & 0.299 & 0.307 \\ 
0010 & -0.099 & -0.107 \\ 
0020 & 0.050 & 0.063 \\ 
0030 & 0.043 & 0.050 \\ 
1000 & -0.099 & -0.107 \\ 
1010 & 0.005 & 0.006 \\ 
1020 & -0.012 & -0.014 \\ 
1030 & -0.002 & -0.003 \\ 
2000 & 0.050 & 0.063 \\ 
2010 & -0.012 & -0.014 \\ 
2020 & 0.008 & 0.012 \\ 
2030 & 0.005 & 0.007 \\ 
3000 & 0.043 & 0.050 \\ 
3010 & -0.002 & -0.003 \\ 
3020 & 0.005 & 0.007 \\ 
3030 & 0.001 & 0.001 \\ 
\hline
\end{tabular}
    \caption{Table of tensor components for numerically checking the crossing relation.}
    \label{tab:crossing}
\end{table}

\begin{table}[H]
    \centering
    \scalebox{0.9}{
        \begin{tabular}{||c|c|c||}
    \hline
    Components & $T_{R.H.S.}$ & $T_{L.H.S.}$ \\ 
    \hline
    0000 & 1.015 & 1.013 \\ 
    0100 & 0.122 & 0.124 \\ 
    0200 & 0.000 & -0.000 \\ 
    0300 & 0.082 & 0.083 \\ 
    1000 & 0.122 & 0.124 \\ 
    1100 & 0.020 & 0.021 \\ 
    1200 & 0.002 & 0.002 \\ 
    1300 & 0.014 & 0.015 \\ 
    2000 & 0.000 & -0.000 \\ 
    2100 & -0.002 & -0.002 \\ 
    2200 & -0.001 & -0.001 \\ 
    2300 & -0.002 & -0.002 \\ 
    3000 & 0.082 & 0.083 \\ 
    3100 & 0.014 & 0.015 \\ 
    3200 & 0.002 & 0.002 \\ 
    3300 & 0.010 & 0.010 \\ 
    \hline
    \end{tabular}
    }
    \caption{Table of tensor components for numerically checking the coarse-graining condition. }
    \label{tab:components}
\end{table}

\section{More details on the transfer matrix}\label{app:transfer}

\begin{figure}[H]
    \centering
    \begin{subfigure}[b]{0.2\textwidth}
         \centering
         \includegraphics[scale=0.35]{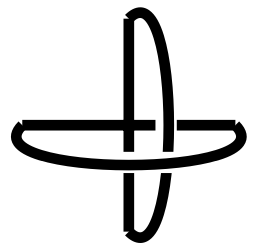}
        \caption{}
        \label{figA8a}
    \end{subfigure}
    \hfill
    \begin{subfigure}[b]{0.25\textwidth}
        \centering
        \includegraphics[scale=0.35]{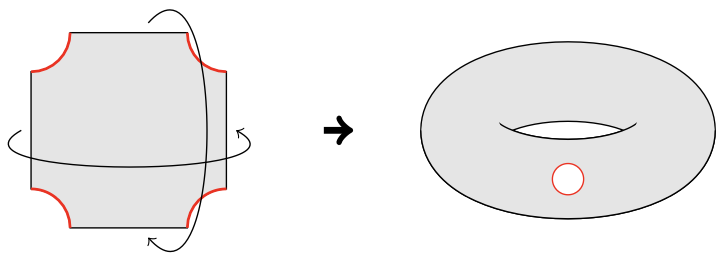}
        \caption{}
        \label{figA8b}
    \end{subfigure}
    \caption{Contraction the opposite legs of fixed point tensor produces partition function on torus.}
    \label{figA8}
\end{figure}

For our construction of FP tensor and the transfer matrix, the eigenvalue reproduce bulk state spectrum is expected because the trace of transfer matrix has a straight foward geometric meaning.  The contraction of opposite legs of FP tensor corresponds to gluing the opposing edges of the square-shaped region, as depicted in \figref{figA8}. The outcome of this procedure is a path integral on a torus with a hole on the surface. As described in the main text, the weighted sum of boundary states residing on this small hole can be projected to the ground state as the size of the hole is reduced to zero.

To obtain the closed string spectrum, we tiling multiple tensors into a cylinder, as shown in \figref{figA10}. For the case where we use $n$-number of rank-4 tensors in the tiling process, the trace computes a torus partition function with moduli $\tau=\frac{1}{n}$, 
\begin{equation}
    \Tr(M_n)=\sum_i e^{\frac{2\pi}{n}(\frac{c}{12}-\Delta_i)}.
\end{equation}

In order for this equation to be true for arbitrary value of $n$, the transfer matrix $M_n$ must be able to diagonalize and give 
rise the following spectrum of CFT bulk states. 
\be
\lambda_{n}(i)=e^{\frac{2\pi}{n}(\frac{c}{12}-\Delta_i)}
\ee 

In the calculation of the FP tensor, a normalization prefactor always appears. However, this can be eliminated by considering the following ratio.
\begin{equation}
    \frac{\lambda_{2}(0)}{\lambda_1(0)^2}=e^{-\frac{\pi}{4}c}.
\end{equation}

This way we obtain the central charge $c$. The value computed from the transfer matrix is listed in the Table \ref{tab:central} and Table \ref{tab:charge}.

In the following, we provide the spectrum data calculated from various cylinder moduli and bond dimensions in \tabref{tab:cyl4}, \ref{tab:cyl3}, \ref{tab:cyl2}. Each column in these tables is labeled according to the cutoff in descendant levels for the fields on the three tensor legs. For example, 447 means that the short edges of the triangle have descendant level cut-off 4 while the long edge has descendant level cut-off 7.   The data clearly demonstrate a consistent convergence of the computed conformal dimension towards the precise value as the cutoff increases.

The central charge can be computed with higher accuracy than  conformal dimensions  In general, lower conformal dimensions are easier to determine than higher ones, as the latter require a larger cutoff in the tensor bond dimension. The main computational challenge arises from obtaining the conformal blocks of descendant fields. 
Our method efficiently extracts the low-lying components of the fixed-point tensor, which can be directly computed using Eq.~\eqref{eq:primarytensor}. However, retrieving higher-order components becomes increasingly difficult, as it requires computing the conformal blocks of descendant fields using the recursive equations presented in the \appref{app:recursive}. These calculations grow increasingly complex at higher descendant levels.

A key advantage of this algorithm is that it provides analytical expressions for the conformal blocks, avoiding the need for recomputation across different CFT models.

\begin{table}
\centering
\begin{tabular}{||c|c|c|c|c|c|c||}
\hline
State/Cutoff & 007    & 117    & 227    & 337    & 447    & exact    \\ 
\hline
$\I$                               & 0.0000  & 0.0000  & 0.0000  & 0.0000  & 0.0000  & 0.0000  \\
$\si$                              & 0.1245  & 0.1249  & 0.1249  & 0.1250  & 0.1250  & 0.1250  \\
$\psi$                         & 1.0293  & 1.0029  & 0.9930  & 0.9963  & 0.9989  & 1.0000  \\
$\partial\si, \bar{\partial}\si$   & 1.1708  & 1.1324  & 1.1255  & 1.1237  & 1.1253  & 1.1250  \\
$\partial\psi, \bar{\partial}\psi$ & 2.1345 & 2.0196 & 2.0030 & 1.9985 & 2.0004  & 2.0000 \\
$\I^{(-2)}$, $\bar{\I}^{(-2)}$     & 2.2234  & 2.0301  & 1.9859  & 1.9899  & 1.9986  & 2.0000  \\
$\partial\bar{\partial}\sigma$     & 2.3037  & 2.1449  & 2.1227  & 2.1045  & 2.1099  & 2.1250  \\
$\si^{(-2)}$, $\bar{\si}^{(-2)}$   & 2.5273  & 2.1591  & 2.1330  & 2.1086  & 2.1208  & 2.1250  \\
$\partial\bar{\partial}\psi$   & 3.0715  & 3.0345  & 2.9904  & 2.9384  & 2.9517  & 3.0000  \\
$\partial^2\psi$, $\bar{\partial}^2\psi$ & 2.9115 & 3.0568 & 2.9732 & 2.9929 & 3.0030 & 3.0000 \\
$\I^{(-3)}, \bar{\I}^{(-3)}$       & 2.9839  & 3.0491  & 2.9849  & 2.9333  & 2.9496  & 3.0000  \\
$\bar{\partial}\si^{(-2)},\partial \bar{\si}^{(-2)}$ & 3.0689 & 3.1085 & 3.0083 & 2.9444 & 2.9900 & 3.1250 \\
$\partial \si^{(-2)}$, $\bar{\partial}\bar{\si}^{(-2)}$ & 3.6670 & 3.1805 & 3.0902 & 3.1065 & 3.1167 & 3.1250 \\
$\si^{(-3)},\bar{\si}^{(-3)}$              & 3.6973  & 3.2351  & 3.1346  & 3.1085  & 3.1291  & 3.1250  \\
$\I^{(-2,-\bar{2})}$                               & 4.0464  & 4.0442  & 4.0446  & 4.0431  & 4.0323  & 4.0000  \\
\hline
\end{tabular}
    \caption{Data from cylinder with length $n=4$. Columns are labeled by the cutoff in descendant levels of the three legs of tensor. The actual bond dimensions correspond to the number of fields up to the specified descendant level cutoff.  For instance, the bond dimensions for a cutoff of 447 is (7,7,22).  }
    \label{tab:cyl4}
\end{table}

\begin{table}
\centering
\begin{tabular}{||c|c|c|c|c|c|c||}
\hline
 State/Cutoff     & 007   & 117   & 227   & 337   & 447  & exact \\ 
\hline
$\I$                              & 0.0000 & 0.0000 & 0.0000 & 0.0000 & 0.0000 & 0.0000 \\
$\si$                             & 0.1241 & 0.1248 & 0.1249 & 0.1250 & 0.1250 & 0.1250 \\
$\psi$                        & 1.0541 & 1.0051 & 0.9878 & 0.9935 & 0.9980  & 1.0000\\
$\partial\si, \bar{\partial}\si$  & 1.2115 & 1.1380 & 1.1258 & 1.1226 & 1.1253 & 1.1250\\
$\partial\psi, \bar{\partial}\psi$ & 2.1398 & 2.0356 & 1.9798 & 1.9873 & 1.9924 & 2.0000 \\
$\I^{(-2)}$, $\bar{\I}^{(-2)}$    & 2.2034 & 2.0554 & 2.0090 & 1.9902 & 2.0038 & 2.0000 \\
$\partial\bar{\partial}\sigma$    & 2.2050 & 2.1618 & 2.0870 & 2.1022 & 2.1204 & 2.1250\\
$\si^{(-2)}$                      & 2.3574 & 2.1582 & 2.0954 & 2.0576 & 2.0737 & 2.1250\\
\hline
\end{tabular}
    \caption{Data from cylinder with length $n=3$. }
    \label{tab:cyl3}
\end{table}

\begin{table}
    \centering
    \begin{tabular}{||c|c|c|c|c|c|c||}
\hline
    State/Cutoff                         & 007   & 117   & 227   & 337   & 447 & exact   \\ \hline
$\I$                         & 0.0000 & 0.0000 & 0.0000 & 0.0000 & 0.0000 & 0.0000 \\
$\si$                        & 0.1231 & 0.1247 & 0.1246 & 0.1250 & 0.1250 & 0.1250\\
$\psi$                   & 1.1146 & 0.9751 & 1.0110 & 0.9862 & 0.9955 & 1.0000\\
$\partial\si, \bar{\partial}\si$ & 1.3039 & 1.1313 & 1.1401 & 1.1166 & 1.1232 & 1.1250 \\
\hline
\end{tabular}
    \caption{Data from cylinder with length $n=2$.}
    \label{tab:cyl2}
\end{table}

\section{Gauge transformation}\label{app:gauge}
To compare our construction of FP tensor with that obtained from TNR method, we have to pick a particular gauge. One convenient choice of the gauge is to rotate the tensor lege to the basis which diagonalizes the transfer matrix $M_{aiI,bjJ}$, 
\begin{equation}
\sum_{a,b,i,I,j,J}M_{aiI,bjJ}O_{aiI}^rO_{bjJ}^s=e^{-2\pi \Delta_r} \delta_{rs}
\end{equation}

Using the same orthogonal matrix $O$ we can rotate the rank-3 tensor to the same basis,
\begin{equation}
\tilde{\mathcal{T}}_{rst}=\sum_{a,i,I,j,J,k,K}\mathcal{T}^{aaa}_{(i,I)(j,J)(k,K)}O_{aiI}^rO_{ajJ}^sO_{akK}^t.
\end{equation}

In Ising CFT, we calculate the rank-3 tensor in diagonal basis and list the leading components in Table~\ref{tab:diag}. 

\begin{table}
    \centering
    \scalebox{0.9}{ 
        \begin{tabular}{||c|c||}
            \hline
            Components & $\tilde{\mathcal{T}}_{rst}$  \\
            \hline
            000 & 1.000  \\
            101 & 0.801 \\
            110 & 0.727  \\
            202 & 0.203  \\
            220 & 0.105  \\
            121 & 0.091  \\
            112 & 0.187 \\
            \hline
        \end{tabular}
    }
    \caption{The rank-3 FP tensor in diagonal basis.}
    \label{tab:diag}
\end{table}

These tensor values can be fit by the three point function of bulk operators. We have normalized the first component to $T_{000}=1$. 

\begin{equation}
   \tilde{\mathcal{T}}_{rst}\approx C_{rst} L^{\Delta_r-\Delta_s-\Delta_t}L^{\Delta_s-\Delta_r-\Delta_t}(\sqrt{2} L)^{\Delta_t-\Delta_r-\Delta_s}
\end{equation}
with $L\approx 2.2$, this matches with the result from numerical TNR \cite{PhysRevResearch.4.023159}\cite{UedaFP}.

\section{Entanglement filtering}\label{app:entfilter}
In the numerical TNR method, besides the SVD decomposition and coarse-graining step there is an additional procedure called  entanglement filtering. Intuitively the purpose of this step is to remove the short range entanglement hidden in the tensor network and thereby get rid of the unphysical components of the tensor. 

\begin{figure}[H]
    \centering
    \includegraphics[scale=0.50]{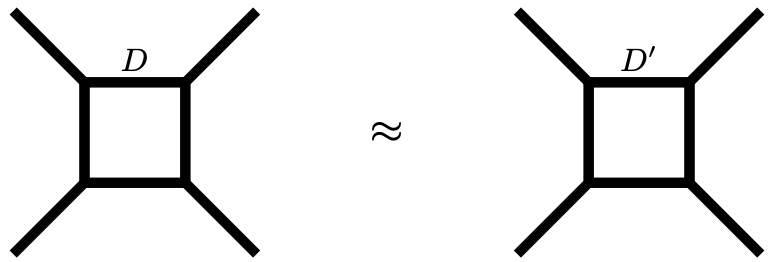}
    \caption{Entanglement filtering procedure. The bond dimension cut-off of internal loop legs being optimized from $D$ to $D'<D$.}
    \label{fig:EEfilter}
\end{figure}

There are several ways to achieve this goal. One of them is plotted in \figref{fig:EEfilter}. This procedure is performed for loops in the tensor network \cite{Gu09}\cite{Yang}. Entanglement filtering corresponds to minimizing the dimension of internal legs within the loop.

Here we comment that the FP tensor we constructed in this work  minimizes the bond dimension cut-off, at least asymptotically. This is because the d.o.f.'s in each leg of FP tensor corresponds to primary and descendant fields of CFT. They are ordered according to their conformal dimensions $h_i$. Since their contribution to the tensor value is proportional to $e^{-h_i}$, the induced error by a conformal dimension cut-off $h_{cut}$ is proportional to $e^{-h_{cut}}$.  A general change of basis in the internal legs changes the ordering of these states and therefore likely to increase the error by keeping the same number of states. 

\begin{figure}[H]
    \centering
    \includegraphics[scale=0.50]{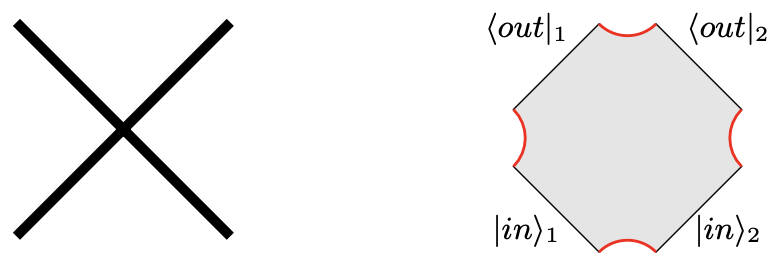}
    \caption{The rank-4 tensor can be viewed as a Euclidean time evolution using the CFT Hamiltonian. }
    \label{fig:positivity}
\end{figure}

Another method in TNR for entanglement filtering is to maintain the positivity of local Hamiltonian in each step of RG \cite{PhysRevLett.118.250602}.  Our FP tensor also satisfies this property, since each rank-4 tensor is computed from the Euclidean path-integral in a square-shaped patch, see \figref{fig:positivity}. It can be naturally viewed as an Euclidean time evolution from the $in$-states to the $out$-states with the CFT Hamiltonian which is positive for unitary models. 


\begin{widetext}

\section{Recursive equation}\label{app:recursive}
\textit{Transformation rules} ---
In this section, we present details in mapping descendant fields under a conformal transformation $\chi(z)$. For more general descendant fields, we can not give a simple expression for the transformation coefficients, but deriving an iteration relation is possible. 

Suppose that we already know the transformation rule for the operator $O^{(-k_l, \cdots,-k_2,-k_1)}:=L_{-k_l}\cdots L_{-k_2}L_{-k_1}O$. The transformation under holomorphic function $\chi(z)$ is written as 
\begin{equation}
    \chi_*\dO{k}(z)=\sum_{\{k'\}\leq\{k\}} \HH{k}{k'}(z)\dO{k'}(\eta) .
\end{equation}
where $\eta=\chi(z)$, and the symbol $\ind{k}$ is a shorthand notation of $\ind{k_l,\dots,k_2,k_1}$. $\ind{k'}\leq\ind{k}$ means that $\forall k_i\in \ind{k}$,  $k'_i\leq k_i$. Moreover, suppose that we also know the OPE between $T(z)$ and $\dO{k}(z')$: 
\begin{equation}
\begin{split}
    T(z)\dO{k}(z')=&\sum_{k'_{l+1}} (z-z')^{k'_{l+1}-2}\ldO{k}{k'}(z')\\
    &+\sum_{\ind{k'}\leq\ind{k}}\frac{\CC{k}{k'}}{(z-z')^{\sum_{p=1}^l k_p-\sum_{p=1}^l k'_p+2}}\dO{k'}(z').
\end{split}
\end{equation}

Then we can derive the transformation rule for higher level descendant field $\ldO{k}{k}$ as 
\begin{equation}
\begin{split}
    &\chi_*\ldO{k}{k}(z')\\[5pt]
    =&\oint_{z'} \dd{z} (z-z')^{1-k_{l+1}} \chi_*T(z) \chi_*\dO{k}(z')\\[5pt]
    =& \oint_{\eta'} \dd{\eta} \frac{(z-z')^{1-k_{l+1}}}{\chi'(z)}[\chi'(z)^2T(\eta)+\frac{c}{12}\{\chi(z),z\}] \sum_{\ind{k'}\leq\ind{k}} \HH{k}{k'}(z')\dO{k'}(\eta')\\[20pt]
    =&\sum_{\ind{k'}\leq\ind{k}}\sum_{k'_{l+1}=0}^{k_{l+1}}\HH{k}{k'}(z')a^{k_{l+1}}_{k_{l+1}-k'_{l+1}}(z')\ldO{k'}{k'}(\eta')\\
    &+\sum_{\ind{k''}\leq\ind{k}}[\sum_{\ind{k''}\leq\ind{k'}\leq\ind{k}}\HH{k}{k'}(z')a^{k_{l+1}}_{k_{l+1}+\sum k'-\sum k''}(z')\CC{k'}{k''}]\dO{k''}(\eta')\\
    &+\frac{c}{12(k_{l+1}-2)!}\left. (\frac{d}{dz})^{k_{l+1}-2}\{\chi(z),z\}\right|_{z=z'}\sum_{\ind{k'}\leq\ind{k}}\HH{k}{k'}(z')\dO{k'}(\eta'),
\end{split}
\end{equation}
where the coefficients $a^n_m$'s are defined by 
\begin{equation}
    \chi'(z)(z-z')^{1-n}=(\eta-\eta')^{1-n}\sum _{m=0}^{\infty}a_m^n(z') (\eta-\eta')^{m}.
\end{equation}

Comparing with the definition of these transformation coefficients we conclude that, 
\begin{equation}
\begin{split}
    \lHH{k}{k'}(z')= &\HH{k}{k'}(z')a^{k_{l+1}}_{k_{l+1}-k'_{l+1}}(z')\\
    \ldHH{k}{k'}(z')=&\sum_{\ind{k'}\leq\ind{k''}\leq\ind{k}}a^{k_{l+1}}_{k_{l+1}+\sum k''-\sum k'}(z') \HH{k}{k''}(z')\CC{k''}{k'}\\
    &+\frac{c}{12(k_{l+1}-2)!}\left. (\frac{d}{dz})^{k_{l+1}-2}\{\chi(z),z\}\right|_{z=z'}\HH{k}{k'}(z').
\end{split}
\end{equation}

\textit{OPE coefficients} ---\label{app42}
Now we derive an iteration relation of the OPE coefficient between energy momentum tensor $T(z)$ and a general descendant field $\dO{k}(z')$. The OPE coefficients are denoted by the symbol $\CC{k}{k'}$ defined as,
\begin{equation}
\begin{split}
    T(z) \dO{k}(z')=&\sum_{k'_{l+1}=1}^{\infty} (z-z')^{k'_{l+1}-2}\ldO{k}{k'}(z')\\&+\sum_{\ind{k'}\leq \ind{k}}\frac{\CC{k}{k'}}{(z-z')^{\sum_{p=1}^l k_p-\sum_{p=1}^l k'_p+2}} \dO{k'}(z').
\end{split}
\end{equation}

Again, we use the symbol $\ind{k}$ to denote the set $\{k_l,\cdots,k_2,k_1\}$. $\ind{k'}\leq \ind{k}$ means that $k'_p\leq k_p$ for any $1\leq p\leq l$. 

For the higher level descendants $\ldO{k}{k}$, its operator product with $T(z)$ is, 
\begin{equation}\label{eqop}
\begin{split}
    &T(z)\ldO{k}{k}(z')\\=&\oint_{z'} \dd{w}(w-z')^{1-k_{l+1}}T(z)T(w)\dO{k}(z')\\
    =&- [\oint_{z} \dd{w} (w-z')^{1-k_{l+1}}T(w) T(z)]\dO{k}(z')+\oint_{z'} \dd{w} (w-z')^{1-k_{l+1}}T(w) [T(z) \dO{k}(z')]\\
    =&\sum_{k'_{l+1}=1}^{k_{l+1}}(2k_{l+1}-k'_{l+1})\frac{ \ldO{k}{k'}}{(z-z')^{k_{l+1}-k'_{l+1}+2}}+ \sum_{\ind{k'}\leq \ind{k}}\CC{k}{k'}\frac{\ldO{k'}{k}(z')}{(z-z')^{\sum_{p=1}^l k_p-\sum_{p=1}^l k'_p+2}}  \\
    &+\sum_{\ind{k'}\leq\ind{k}}(2k_{l+1}+\sum_{p=1}^l k_p-\sum_{p=1}^l k_p')\CC{k}{k'}\frac{\dO{k'}(z')}{(z-z')^{k_{l+1}+\sum_{p=1}^l k_p-\sum_{p=1}^l k_p'+2}}\\
    &+\frac{c}{12}k_{l+1}(k_{l+1}^2-1)\frac{\dO{k}}{(z-z')^{k_{l+1}+2}}+\cdots
\end{split}
\end{equation}

where the ellipsis denotes any combination of descendant operators at level higher than $\sum_{p=1}^{l+1}k_p$. We neglected them simply because we already know their coefficients.

Comparing with the definition of $\lCC{k}{k'}$, we conclude that,
\begin{equation}
    \begin{split}
        &\lsCC{k}{k}{k'}=2k_{l+1}-k'_{l+1},\qquad \text{for} \ 1\leq k'_{l+1}<k_{l+1} \\
        &\lsCC{k}{k'}{k}=\CC{k}{k'}, \qquad \text{for} \ \ind{k'}<\ind{k}\\
        &\lCC{k}{k}=k_{l+1}+\CC{k}{k}\\
        &\ldCC{k}{k'}=(2k_{l+1}+\sum_{p=1}^l k_p-\sum_{p=1}^{l} k'_p)\CC{k}{k'}+\frac{c}{12}k_{l+1}(k_{l+1}^2-1)\delta^{\ind{k}}_{\ind{k'}} \qquad\\
    \end{split}
\end{equation}

\textit{Correlation function} ---\label{app43}
We can also derive an iteration equation of correlators. Suppose that we already know all the correlators of lower level descendants, we can derive the higher level ones using this equation. 

To simplify the notation, we use $O^{(-\vec{k})}$ to denote $\dO{k}$, and the OPE coefficient $C^{\vec{k}}_{\vec{k'}}$ to denote $\CC{k}{k'}$.   Then we will show that the correlator $\langle L_{-m}O_1^{(-\vec{k})}(x)  O_2^{(-\vec{p})}(y)O_3^{(-\vec{q})}(z)\rangle$ can be written as a linear combination of simpler correlators of the form $\langle O_1^{(-\vec{k'})}(x)  O_2^{(-\vec{p'})}(y)O_3^{(-\vec{q'})}(z)\rangle$, with $\vec{k'}\leq \vec{k}$, $\vec{p'}\leq \vec{p}$ and $\vec{q'}\leq \vec{q}$ (in the sense of $\ind{k'} \leq \ind{k}$ defined in the previous sections). We start from the equation, 

\begin{equation}
\begin{split}
    &\langle L_{-m}O_1^{(-\vec{k})}(x)  O_2^{(-\vec{p})}(y)O_3^{(-\vec{q})}(z)\rangle\\
    =&\oint_x \dd{w} (w-x)^{1-m}\langle [T(w)O_1^{(-\vec{k})}(x)]  O_2^{(-\vec{p})}(y)O_3^{(-\vec{q})}(z)\rangle\\
    =& -\oint_y \dd{w} (w-x)^{1-m}\langle \vO{1}{k}(x) [T(w)\vO{2}{p}(y)]\vO{3}{q}(z)\rangle -\oint_z \dd{w} (w-x)^{1-m} \langle \vO{1}{k}(x)\vO{2}{p}(y)[T(w)\vO{3}{q}(z)]\rangle\\
\end{split}
\end{equation}

Using the OPE,
\begin{equation}
    T(w)\vO{}{p}(y)=\sum_{\vec{p'}\leq \vec{p}}C^{\vec{p}}_{\vec{p'}}\frac{\vO{}{p'}(y)}{(w-y)^{|\vec{p}|-|\vec{p'}|+2}}+\frac{\partial \vO{}{p}(y)}{w-y}+reg.,
\end{equation}
where $reg.$ means the regular terms in the limit $w\rightarrow y$, we can expand the expression, 
\begin{equation}
    \begin{split}
       & \oint_y \dd{w} (w-x)^{1-m}\langle \vO{1}{k}(x) [T(w)\vO{2}{p}(y)]\vO{3}{q}(z)\rangle\\
        =& \frac{\partial_y}{(y-x)^{m-1}}\langle \vO{1}{k}(x)\vO{2}{p}(y)\vO{3}{q}(z)\rangle\\
        &+(-1)^{|\vec{p}|-|\vec{p'}|-1} \sum_{\vec{p'}\leq\vec{p}}\frac{\vC{p}{p'}}{(y-x)^{m+|\vec{p}|-|\vec{p'}|}}\frac{(|\vec{p}|-|\vec{p'}|+m-1)!}{(|\vec{p}|-|\vec{p'}|+1)!(m-2)!}\langle \vO{1}{k}(x)\vO{2}{p'}(y)\vO{3}{q}(z)\rangle.
    \end{split}
    \end{equation}

The other term is calculated similarly. So the correlator is reduced to combinations of simpler ones:
\begin{equation}
    \begin{split}
        &\langle L_{-m}O_1^{(-\vec{k})}(x)  O_2^{(-\vec{p})}(y)O_3^{(-\vec{q})}(z)\rangle\\
        =& -\frac{\partial_y}{(y-x)^{m-1}}\langle \vO{1}{k}(x)\vO{2}{p}(y)\vO{3}{q}(z)\rangle\\
        &+(-1)^{|\vec{p}|-|\vec{p'}|} \sum_{\vec{p'}\leq\vec{p}}\frac{\vC{p}{p'}}{(y-x)^{m+|\vec{p}|-|\vec{p'}|}}\frac{(|\vec{p}|-|\vec{p'}|+m-1)!}{(|\vec{p}|-|\vec{p'}|+1)!(m-2)!}\langle \vO{1}{k}(x)\vO{2}{p'}(y)\vO{3}{q}(z)\rangle\\
        &+(y\rightarrow z, \ \vec{p}\rightarrow \vec{q}).
    \end{split}
\end{equation}
\end{widetext}

\bibliography{ref}

\begin{thebibliography}{60}%
\makeatletter
\providecommand \@ifxundefined [1]{%
 \@ifx{#1\undefined}
}%
\providecommand \@ifnum [1]{%
 \ifnum #1\expandafter \@firstoftwo
 \else \expandafter \@secondoftwo
 \fi
}%
\providecommand \@ifx [1]{%
 \ifx #1\expandafter \@firstoftwo
 \else \expandafter \@secondoftwo
 \fi
}%
\providecommand \natexlab [1]{#1}%
\providecommand \enquote  [1]{``#1''}%
\providecommand \bibnamefont  [1]{#1}%
\providecommand \bibfnamefont [1]{#1}%
\providecommand \citenamefont [1]{#1}%
\providecommand \href@noop [0]{\@secondoftwo}%
\providecommand \href [0]{\begingroup \@sanitize@url \@href}%
\providecommand \@href[1]{\@@startlink{#1}\@@href}%
\providecommand \@@href[1]{\endgroup#1\@@endlink}%
\providecommand \@sanitize@url [0]{\catcode `\\12\catcode `\$12\catcode
  `\&12\catcode `\#12\catcode `\^12\catcode `\_12\catcode `\%12\relax}%
\providecommand \@@startlink[1]{}%
\providecommand \@@endlink[0]{}%
\providecommand \url  [0]{\begingroup\@sanitize@url \@url }%
\providecommand \@url [1]{\endgroup\@href {#1}{\urlprefix }}%
\providecommand \urlprefix  [0]{URL }%
\providecommand \Eprint [0]{\href }%
\providecommand \doibase [0]{http://dx.doi.org/}%
\providecommand \selectlanguage [0]{\@gobble}%
\providecommand \bibinfo  [0]{\@secondoftwo}%
\providecommand \bibfield  [0]{\@secondoftwo}%
\providecommand \translation [1]{[#1]}%
\providecommand \BibitemOpen [0]{}%
\providecommand \bibitemStop [0]{}%
\providecommand \bibitemNoStop [0]{.\EOS\space}%
\providecommand \EOS [0]{\spacefactor3000\relax}%
\providecommand \BibitemShut  [1]{\csname bibitem#1\endcsname}%
\let\auto@bib@innerbib\@empty
\bibitem [{\citenamefont {Vidal}(2007)}]{VidalER}%
  \BibitemOpen
  \bibfield  {author} {\bibinfo {author} {\bibfnamefont {G.}~\bibnamefont
  {Vidal}},\ }\bibfield  {title} {\enquote {\bibinfo {title} {Entanglement
  renormalization},}\ }\href {\doibase 10.1103/PhysRevLett.99.220405}
  {\bibfield  {journal} {\bibinfo  {journal} {Phys. Rev. Lett.}\ }\textbf
  {\bibinfo {volume} {99}},\ \bibinfo {pages} {220405} (\bibinfo {year}
  {2007})}\BibitemShut {NoStop}%
\bibitem [{\citenamefont {Gu}\ \emph {et~al.}(2008)\citenamefont {Gu},
  \citenamefont {Levin},\ and\ \citenamefont {Wen}}]{Gu08}%
  \BibitemOpen
  \bibfield  {author} {\bibinfo {author} {\bibfnamefont {Zheng-Cheng}\
  \bibnamefont {Gu}}, \bibinfo {author} {\bibfnamefont {Michael}\ \bibnamefont
  {Levin}}, \ and\ \bibinfo {author} {\bibfnamefont {Xiao-Gang}\ \bibnamefont
  {Wen}},\ }\bibfield  {title} {\enquote {\bibinfo {title} {Tensor-entanglement
  renormalization group approach as a unified method for symmetry breaking and
  topological phase transitions},}\ }\href {\doibase
  10.1103/PhysRevB.78.205116} {\bibfield  {journal} {\bibinfo  {journal} {Phys.
  Rev. B}\ }\textbf {\bibinfo {volume} {78}},\ \bibinfo {pages} {205116}
  (\bibinfo {year} {2008})}\BibitemShut {NoStop}%
\bibitem [{\citenamefont {Evenbly}\ and\ \citenamefont
  {Vidal}(2009{\natexlab{a}})}]{Vidal09a}%
  \BibitemOpen
  \bibfield  {author} {\bibinfo {author} {\bibfnamefont {G.}~\bibnamefont
  {Evenbly}}\ and\ \bibinfo {author} {\bibfnamefont {G.}~\bibnamefont
  {Vidal}},\ }\bibfield  {title} {\enquote {\bibinfo {title} {Algorithms for
  entanglement renormalization},}\ }\href {\doibase 10.1103/PhysRevB.79.144108}
  {\bibfield  {journal} {\bibinfo  {journal} {Phys. Rev. B}\ }\textbf {\bibinfo
  {volume} {79}},\ \bibinfo {pages} {144108} (\bibinfo {year}
  {2009}{\natexlab{a}})}\BibitemShut {NoStop}%
\bibitem [{\citenamefont {Pfeifer}\ \emph {et~al.}(2009)\citenamefont
  {Pfeifer}, \citenamefont {Evenbly},\ and\ \citenamefont {Vidal}}]{Vidal09b}%
  \BibitemOpen
  \bibfield  {author} {\bibinfo {author} {\bibfnamefont {Robert N.~C.}\
  \bibnamefont {Pfeifer}}, \bibinfo {author} {\bibfnamefont {Glen}\
  \bibnamefont {Evenbly}}, \ and\ \bibinfo {author} {\bibfnamefont {Guifr\'e}\
  \bibnamefont {Vidal}},\ }\bibfield  {title} {\enquote {\bibinfo {title}
  {Entanglement renormalization, scale invariance, and quantum criticality},}\
  }\href {\doibase 10.1103/PhysRevA.79.040301} {\bibfield  {journal} {\bibinfo
  {journal} {Phys. Rev. A}\ }\textbf {\bibinfo {volume} {79}},\ \bibinfo
  {pages} {040301} (\bibinfo {year} {2009})}\BibitemShut {NoStop}%
\bibitem [{\citenamefont {Gu}\ and\ \citenamefont {Wen}(2009)}]{Gu09}%
  \BibitemOpen
  \bibfield  {author} {\bibinfo {author} {\bibfnamefont {Zheng-Cheng}\
  \bibnamefont {Gu}}\ and\ \bibinfo {author} {\bibfnamefont {Xiao-Gang}\
  \bibnamefont {Wen}},\ }\bibfield  {title} {\enquote {\bibinfo {title}
  {Tensor-entanglement-filtering renormalization approach and
  symmetry-protected topological order},}\ }\href {\doibase
  10.1103/PhysRevB.80.155131} {\bibfield  {journal} {\bibinfo  {journal} {Phys.
  Rev. B}\ }\textbf {\bibinfo {volume} {80}},\ \bibinfo {pages} {155131}
  (\bibinfo {year} {2009})}\BibitemShut {NoStop}%
\bibitem [{\citenamefont {Evenbly}\ and\ \citenamefont
  {Vidal}(2009{\natexlab{b}})}]{PhysRevLett.102.180406}%
  \BibitemOpen
  \bibfield  {author} {\bibinfo {author} {\bibfnamefont {G.}~\bibnamefont
  {Evenbly}}\ and\ \bibinfo {author} {\bibfnamefont {G.}~\bibnamefont
  {Vidal}},\ }\bibfield  {title} {\enquote {\bibinfo {title} {Entanglement
  renormalization in two spatial dimensions},}\ }\href {\doibase
  10.1103/PhysRevLett.102.180406} {\bibfield  {journal} {\bibinfo  {journal}
  {Phys. Rev. Lett.}\ }\textbf {\bibinfo {volume} {102}},\ \bibinfo {pages}
  {180406} (\bibinfo {year} {2009}{\natexlab{b}})}\BibitemShut {NoStop}%
\bibitem [{\citenamefont {Levin}\ and\ \citenamefont {Nave}(2007)}]{Levin}%
  \BibitemOpen
  \bibfield  {author} {\bibinfo {author} {\bibfnamefont {Michael}\ \bibnamefont
  {Levin}}\ and\ \bibinfo {author} {\bibfnamefont {Cody~P.}\ \bibnamefont
  {Nave}},\ }\bibfield  {title} {\enquote {\bibinfo {title} {Tensor
  renormalization group approach to two-dimensional classical lattice
  models},}\ }\href {\doibase 10.1103/PhysRevLett.99.120601} {\bibfield
  {journal} {\bibinfo  {journal} {Phys. Rev. Lett.}\ }\textbf {\bibinfo
  {volume} {99}},\ \bibinfo {pages} {120601} (\bibinfo {year}
  {2007})}\BibitemShut {NoStop}%
\bibitem [{\citenamefont {Xie}\ \emph {et~al.}(2009)\citenamefont {Xie},
  \citenamefont {Jiang}, \citenamefont {Chen}, \citenamefont {Weng},\ and\
  \citenamefont {Xiang}}]{Xie}%
  \BibitemOpen
  \bibfield  {author} {\bibinfo {author} {\bibfnamefont {Z.~Y.}\ \bibnamefont
  {Xie}}, \bibinfo {author} {\bibfnamefont {H.~C.}\ \bibnamefont {Jiang}},
  \bibinfo {author} {\bibfnamefont {Q.~N.}\ \bibnamefont {Chen}}, \bibinfo
  {author} {\bibfnamefont {Z.~Y.}\ \bibnamefont {Weng}}, \ and\ \bibinfo
  {author} {\bibfnamefont {T.}~\bibnamefont {Xiang}},\ }\bibfield  {title}
  {\enquote {\bibinfo {title} {Second renormalization of tensor-network
  states},}\ }\href {\doibase 10.1103/PhysRevLett.103.160601} {\bibfield
  {journal} {\bibinfo  {journal} {Phys. Rev. Lett.}\ }\textbf {\bibinfo
  {volume} {103}},\ \bibinfo {pages} {160601} (\bibinfo {year}
  {2009})}\BibitemShut {NoStop}%
\bibitem [{\citenamefont {Xie}\ \emph {et~al.}(2012)\citenamefont {Xie},
  \citenamefont {Chen}, \citenamefont {Qin}, \citenamefont {Zhu}, \citenamefont
  {Yang},\ and\ \citenamefont {Xiang}}]{HOTRG}%
  \BibitemOpen
  \bibfield  {author} {\bibinfo {author} {\bibfnamefont {Z.~Y.}\ \bibnamefont
  {Xie}}, \bibinfo {author} {\bibfnamefont {J.}~\bibnamefont {Chen}}, \bibinfo
  {author} {\bibfnamefont {M.~P.}\ \bibnamefont {Qin}}, \bibinfo {author}
  {\bibfnamefont {J.~W.}\ \bibnamefont {Zhu}}, \bibinfo {author} {\bibfnamefont
  {L.~P.}\ \bibnamefont {Yang}}, \ and\ \bibinfo {author} {\bibfnamefont
  {T.}~\bibnamefont {Xiang}},\ }\bibfield  {title} {\enquote {\bibinfo {title}
  {Coarse-graining renormalization by higher-order singular value
  decomposition},}\ }\href {\doibase 10.1103/PhysRevB.86.045139} {\bibfield
  {journal} {\bibinfo  {journal} {Phys. Rev. B}\ }\textbf {\bibinfo {volume}
  {86}},\ \bibinfo {pages} {045139} (\bibinfo {year} {2012})}\BibitemShut
  {NoStop}%
\bibitem [{\citenamefont {Evenbly}\ and\ \citenamefont
  {Vidal}(2015)}]{VidalTNR}%
  \BibitemOpen
  \bibfield  {author} {\bibinfo {author} {\bibfnamefont {G.}~\bibnamefont
  {Evenbly}}\ and\ \bibinfo {author} {\bibfnamefont {G.}~\bibnamefont
  {Vidal}},\ }\bibfield  {title} {\enquote {\bibinfo {title} {Tensor network
  renormalization},}\ }\href {\doibase 10.1103/PhysRevLett.115.180405}
  {\bibfield  {journal} {\bibinfo  {journal} {Phys. Rev. Lett.}\ }\textbf
  {\bibinfo {volume} {115}},\ \bibinfo {pages} {180405} (\bibinfo {year}
  {2015})}\BibitemShut {NoStop}%
\bibitem [{\citenamefont {Yang}\ \emph {et~al.}(2017)\citenamefont {Yang},
  \citenamefont {Gu},\ and\ \citenamefont {Wen}}]{Yang}%
  \BibitemOpen
  \bibfield  {author} {\bibinfo {author} {\bibfnamefont {Shuo}\ \bibnamefont
  {Yang}}, \bibinfo {author} {\bibfnamefont {Zheng-Cheng}\ \bibnamefont {Gu}},
  \ and\ \bibinfo {author} {\bibfnamefont {Xiao-Gang}\ \bibnamefont {Wen}},\
  }\bibfield  {title} {\enquote {\bibinfo {title} {Loop optimization for tensor
  network renormalization},}\ }\href {\doibase 10.1103/PhysRevLett.118.110504}
  {\bibfield  {journal} {\bibinfo  {journal} {Phys. Rev. Lett.}\ }\textbf
  {\bibinfo {volume} {118}},\ \bibinfo {pages} {110504} (\bibinfo {year}
  {2017})}\BibitemShut {NoStop}%
\bibitem [{\citenamefont {Evenbly}(2017)}]{PhysRevB.95.045117}%
  \BibitemOpen
  \bibfield  {author} {\bibinfo {author} {\bibfnamefont {G.}~\bibnamefont
  {Evenbly}},\ }\bibfield  {title} {\enquote {\bibinfo {title} {Algorithms for
  tensor network renormalization},}\ }\href {\doibase
  10.1103/PhysRevB.95.045117} {\bibfield  {journal} {\bibinfo  {journal} {Phys.
  Rev. B}\ }\textbf {\bibinfo {volume} {95}},\ \bibinfo {pages} {045117}
  (\bibinfo {year} {2017})}\BibitemShut {NoStop}%
\bibitem [{\citenamefont {Hauru}\ \emph {et~al.}(2016)\citenamefont {Hauru},
  \citenamefont {Evenbly}, \citenamefont {Ho}, \citenamefont {Gaiotto},\ and\
  \citenamefont {Vidal}}]{PhysRevB.94.115125}%
  \BibitemOpen
  \bibfield  {author} {\bibinfo {author} {\bibfnamefont {Markus}\ \bibnamefont
  {Hauru}}, \bibinfo {author} {\bibfnamefont {Glen}\ \bibnamefont {Evenbly}},
  \bibinfo {author} {\bibfnamefont {Wen~Wei}\ \bibnamefont {Ho}}, \bibinfo
  {author} {\bibfnamefont {Davide}\ \bibnamefont {Gaiotto}}, \ and\ \bibinfo
  {author} {\bibfnamefont {Guifre}\ \bibnamefont {Vidal}},\ }\bibfield  {title}
  {\enquote {\bibinfo {title} {Topological conformal defects with tensor
  networks},}\ }\href {\doibase 10.1103/PhysRevB.94.115125} {\bibfield
  {journal} {\bibinfo  {journal} {Phys. Rev. B}\ }\textbf {\bibinfo {volume}
  {94}},\ \bibinfo {pages} {115125} (\bibinfo {year} {2016})}\BibitemShut
  {NoStop}%
\bibitem [{\citenamefont {Iino}\ \emph {et~al.}(2019)\citenamefont {Iino},
  \citenamefont {Morita},\ and\ \citenamefont
  {Kawashima}}]{PhysRevB.100.035449}%
  \BibitemOpen
  \bibfield  {author} {\bibinfo {author} {\bibfnamefont {Shumpei}\ \bibnamefont
  {Iino}}, \bibinfo {author} {\bibfnamefont {Satoshi}\ \bibnamefont {Morita}},
  \ and\ \bibinfo {author} {\bibfnamefont {Naoki}\ \bibnamefont {Kawashima}},\
  }\bibfield  {title} {\enquote {\bibinfo {title} {Boundary tensor
  renormalization group},}\ }\href {\doibase 10.1103/PhysRevB.100.035449}
  {\bibfield  {journal} {\bibinfo  {journal} {Phys. Rev. B}\ }\textbf {\bibinfo
  {volume} {100}},\ \bibinfo {pages} {035449} (\bibinfo {year}
  {2019})}\BibitemShut {NoStop}%
\bibitem [{\citenamefont {Evenbly}\ \emph {et~al.}(2010)\citenamefont
  {Evenbly}, \citenamefont {Pfeifer}, \citenamefont {Pic\'o}, \citenamefont
  {Iblisdir}, \citenamefont {Tagliacozzo}, \citenamefont {McCulloch},\ and\
  \citenamefont {Vidal}}]{PhysRevB.82.161107}%
  \BibitemOpen
  \bibfield  {author} {\bibinfo {author} {\bibfnamefont {G.}~\bibnamefont
  {Evenbly}}, \bibinfo {author} {\bibfnamefont {R.~N.~C.}\ \bibnamefont
  {Pfeifer}}, \bibinfo {author} {\bibfnamefont {V.}~\bibnamefont {Pic\'o}},
  \bibinfo {author} {\bibfnamefont {S.}~\bibnamefont {Iblisdir}}, \bibinfo
  {author} {\bibfnamefont {L.}~\bibnamefont {Tagliacozzo}}, \bibinfo {author}
  {\bibfnamefont {I.~P.}\ \bibnamefont {McCulloch}}, \ and\ \bibinfo {author}
  {\bibfnamefont {G.}~\bibnamefont {Vidal}},\ }\bibfield  {title} {\enquote
  {\bibinfo {title} {Boundary quantum critical phenomena with entanglement
  renormalization},}\ }\href {\doibase 10.1103/PhysRevB.82.161107} {\bibfield
  {journal} {\bibinfo  {journal} {Phys. Rev. B}\ }\textbf {\bibinfo {volume}
  {82}},\ \bibinfo {pages} {161107} (\bibinfo {year} {2010})}\BibitemShut
  {NoStop}%
\bibitem [{\citenamefont {Li}\ \emph {et~al.}(2022)\citenamefont {Li},
  \citenamefont {Pai},\ and\ \citenamefont {Gu}}]{PhysRevResearch.4.023159}%
  \BibitemOpen
  \bibfield  {author} {\bibinfo {author} {\bibfnamefont {Guanrong}\
  \bibnamefont {Li}}, \bibinfo {author} {\bibfnamefont {Kwok~Ho}\ \bibnamefont
  {Pai}}, \ and\ \bibinfo {author} {\bibfnamefont {Zheng-Cheng}\ \bibnamefont
  {Gu}},\ }\bibfield  {title} {\enquote {\bibinfo {title} {Tensor-network
  renormalization approach to the $q$-state clock model},}\ }\href {\doibase
  10.1103/PhysRevResearch.4.023159} {\bibfield  {journal} {\bibinfo  {journal}
  {Phys. Rev. Res.}\ }\textbf {\bibinfo {volume} {4}},\ \bibinfo {pages}
  {023159} (\bibinfo {year} {2022})}\BibitemShut {NoStop}%
\bibitem [{\citenamefont {Ueda}\ and\ \citenamefont {Yamazaki}(2023)}]{UedaFP}%
  \BibitemOpen
  \bibfield  {author} {\bibinfo {author} {\bibfnamefont {Atsushi}\ \bibnamefont
  {Ueda}}\ and\ \bibinfo {author} {\bibfnamefont {Masahito}\ \bibnamefont
  {Yamazaki}},\ }\bibfield  {title} {\enquote {\bibinfo {title} {Fixed-point
  tensor is a four-point function},}\ }\href@noop {} {\  (\bibinfo {year}
  {2023})},\ \Eprint {http://arxiv.org/abs/2307.02523} {arXiv:2307.02523
  [cond-mat.stat-mech]} \BibitemShut {NoStop}%
\bibitem [{\citenamefont {Ji}\ and\ \citenamefont {Wen}(2020)}]{Ji:2019jhk}%
  \BibitemOpen
  \bibfield  {author} {\bibinfo {author} {\bibfnamefont {Wenjie}\ \bibnamefont
  {Ji}}\ and\ \bibinfo {author} {\bibfnamefont {Xiao-Gang}\ \bibnamefont
  {Wen}},\ }\bibfield  {title} {\enquote {\bibinfo {title} {{Categorical
  symmetry and noninvertible anomaly in symmetry-breaking and topological phase
  transitions}},}\ }\href {\doibase 10.1103/PhysRevResearch.2.033417}
  {\bibfield  {journal} {\bibinfo  {journal} {Phys. Rev. Res.}\ }\textbf
  {\bibinfo {volume} {2}},\ \bibinfo {pages} {033417} (\bibinfo {year}
  {2020})},\ \Eprint {http://arxiv.org/abs/1912.13492} {arXiv:1912.13492
  [cond-mat.str-el]} \BibitemShut {NoStop}%
\bibitem [{\citenamefont {Kong}\ \emph
  {et~al.}(2020{\natexlab{a}})\citenamefont {Kong}, \citenamefont {Lan},
  \citenamefont {Wen}, \citenamefont {Zhang},\ and\ \citenamefont
  {Zheng}}]{Kong:2020cie}%
  \BibitemOpen
  \bibfield  {author} {\bibinfo {author} {\bibfnamefont {Liang}\ \bibnamefont
  {Kong}}, \bibinfo {author} {\bibfnamefont {Tian}\ \bibnamefont {Lan}},
  \bibinfo {author} {\bibfnamefont {Xiao-Gang}\ \bibnamefont {Wen}}, \bibinfo
  {author} {\bibfnamefont {Zhi-Hao}\ \bibnamefont {Zhang}}, \ and\ \bibinfo
  {author} {\bibfnamefont {Hao}\ \bibnamefont {Zheng}},\ }\bibfield  {title}
  {\enquote {\bibinfo {title} {{Algebraic higher symmetry and categorical
  symmetry -- a holographic and entanglement view of symmetry}},}\ }\href
  {\doibase 10.1103/PhysRevResearch.2.043086} {\bibfield  {journal} {\bibinfo
  {journal} {Phys. Rev. Res.}\ }\textbf {\bibinfo {volume} {2}},\ \bibinfo
  {pages} {043086} (\bibinfo {year} {2020}{\natexlab{a}})},\ \Eprint
  {http://arxiv.org/abs/2005.14178} {arXiv:2005.14178 [cond-mat.str-el]}
  \BibitemShut {NoStop}%
\bibitem [{\citenamefont {Chatterjee}\ and\ \citenamefont
  {Wen}(2023{\natexlab{a}})}]{catc}%
  \BibitemOpen
  \bibfield  {author} {\bibinfo {author} {\bibfnamefont {Arkya}\ \bibnamefont
  {Chatterjee}}\ and\ \bibinfo {author} {\bibfnamefont {Xiao-Gang}\
  \bibnamefont {Wen}},\ }\bibfield  {title} {\enquote {\bibinfo {title}
  {Holographic theory for continuous phase transitions: Emergence and symmetry
  protection of gaplessness},}\ }\href {\doibase 10.1103/PhysRevB.108.075105}
  {\bibfield  {journal} {\bibinfo  {journal} {Phys. Rev. B}\ }\textbf {\bibinfo
  {volume} {108}},\ \bibinfo {pages} {075105} (\bibinfo {year}
  {2023}{\natexlab{a}})}\BibitemShut {NoStop}%
\bibitem [{\citenamefont {Hung}\ and\ \citenamefont
  {Wong}(2021)}]{Hung:2019bnq}%
  \BibitemOpen
  \bibfield  {author} {\bibinfo {author} {\bibfnamefont {Ling~Yan}\
  \bibnamefont {Hung}}\ and\ \bibinfo {author} {\bibfnamefont {Gabriel}\
  \bibnamefont {Wong}},\ }\bibfield  {title} {\enquote {\bibinfo {title}
  {{Entanglement branes and factorization in conformal field theory}},}\ }\href
  {\doibase 10.1103/PhysRevD.104.026012} {\bibfield  {journal} {\bibinfo
  {journal} {Phys. Rev. D}\ }\textbf {\bibinfo {volume} {104}},\ \bibinfo
  {pages} {026012} (\bibinfo {year} {2021})},\ \Eprint
  {http://arxiv.org/abs/1912.11201} {arXiv:1912.11201 [hep-th]} \BibitemShut
  {NoStop}%
\bibitem [{\citenamefont {Brehm}\ and\ \citenamefont
  {Runkel}(2022)}]{Brehm:2021wev}%
  \BibitemOpen
  \bibfield  {author} {\bibinfo {author} {\bibfnamefont {Enrico~M.}\
  \bibnamefont {Brehm}}\ and\ \bibinfo {author} {\bibfnamefont {Ingo}\
  \bibnamefont {Runkel}},\ }\bibfield  {title} {\enquote {\bibinfo {title}
  {{Lattice models from CFT on surfaces with holes: I. Torus partition function
  via two lattice cells}},}\ }\href {\doibase 10.1088/1751-8121/ac6a91}
  {\bibfield  {journal} {\bibinfo  {journal} {J. Phys. A}\ }\textbf {\bibinfo
  {volume} {55}},\ \bibinfo {pages} {235001} (\bibinfo {year} {2022})},\
  \Eprint {http://arxiv.org/abs/2112.01563} {arXiv:2112.01563
  [cond-mat.stat-mech]} \BibitemShut {NoStop}%
\bibitem [{\citenamefont {Gaiotto}\ and\ \citenamefont
  {Kulp}(2021)}]{Gaiotto:2020iye}%
  \BibitemOpen
  \bibfield  {author} {\bibinfo {author} {\bibfnamefont {Davide}\ \bibnamefont
  {Gaiotto}}\ and\ \bibinfo {author} {\bibfnamefont {Justin}\ \bibnamefont
  {Kulp}},\ }\bibfield  {title} {\enquote {\bibinfo {title} {{Orbifold
  groupoids}},}\ }\href {\doibase 10.1007/JHEP02(2021)132} {\bibfield
  {journal} {\bibinfo  {journal} {JHEP}\ }\textbf {\bibinfo {volume} {02}},\
  \bibinfo {pages} {132} (\bibinfo {year} {2021})},\ \Eprint
  {http://arxiv.org/abs/2008.05960} {arXiv:2008.05960 [hep-th]} \BibitemShut
  {NoStop}%
\bibitem [{\citenamefont {Freed}\ \emph {et~al.}(2022)\citenamefont {Freed},
  \citenamefont {Moore},\ and\ \citenamefont {Teleman}}]{Freed:2022qnc}%
  \BibitemOpen
  \bibfield  {author} {\bibinfo {author} {\bibfnamefont {Daniel~S.}\
  \bibnamefont {Freed}}, \bibinfo {author} {\bibfnamefont {Gregory~W.}\
  \bibnamefont {Moore}}, \ and\ \bibinfo {author} {\bibfnamefont {Constantin}\
  \bibnamefont {Teleman}},\ }\bibfield  {title} {\enquote {\bibinfo {title}
  {{Topological symmetry in quantum field theory}},}\ }\href@noop {} {\
  (\bibinfo {year} {2022})},\ \Eprint {http://arxiv.org/abs/2209.07471}
  {arXiv:2209.07471 [hep-th]} \BibitemShut {NoStop}%
\bibitem [{\citenamefont {Kong}\ \emph
  {et~al.}(2020{\natexlab{b}})\citenamefont {Kong}, \citenamefont {Lan},
  \citenamefont {Wen}, \citenamefont {Zhang},\ and\ \citenamefont
  {Zheng}}]{Kong:2020jne}%
  \BibitemOpen
  \bibfield  {author} {\bibinfo {author} {\bibfnamefont {Liang}\ \bibnamefont
  {Kong}}, \bibinfo {author} {\bibfnamefont {Tian}\ \bibnamefont {Lan}},
  \bibinfo {author} {\bibfnamefont {Xiao-Gang}\ \bibnamefont {Wen}}, \bibinfo
  {author} {\bibfnamefont {Zhi-Hao}\ \bibnamefont {Zhang}}, \ and\ \bibinfo
  {author} {\bibfnamefont {Hao}\ \bibnamefont {Zheng}},\ }\bibfield  {title}
  {\enquote {\bibinfo {title} {{Classification of topological phases with
  finite internal symmetries in all dimensions}},}\ }\href {\doibase
  10.1007/JHEP09(2020)093} {\bibfield  {journal} {\bibinfo  {journal} {JHEP}\
  }\textbf {\bibinfo {volume} {09}},\ \bibinfo {pages} {093} (\bibinfo {year}
  {2020}{\natexlab{b}})},\ \Eprint {http://arxiv.org/abs/2003.08898}
  {arXiv:2003.08898 [math-ph]} \BibitemShut {NoStop}%
\bibitem [{\citenamefont {Apruzzi}\ \emph {et~al.}(2023)\citenamefont
  {Apruzzi}, \citenamefont {Bonetti}, \citenamefont {Garc\'\i{}a~Etxebarria},
  \citenamefont {Hosseini},\ and\ \citenamefont
  {Schafer-Nameki}}]{Apruzzi:2021nmk}%
  \BibitemOpen
  \bibfield  {author} {\bibinfo {author} {\bibfnamefont {Fabio}\ \bibnamefont
  {Apruzzi}}, \bibinfo {author} {\bibfnamefont {Federico}\ \bibnamefont
  {Bonetti}}, \bibinfo {author} {\bibfnamefont {I\~naki}\ \bibnamefont
  {Garc\'\i{}a~Etxebarria}}, \bibinfo {author} {\bibfnamefont {Saghar~S.}\
  \bibnamefont {Hosseini}}, \ and\ \bibinfo {author} {\bibfnamefont {Sakura}\
  \bibnamefont {Schafer-Nameki}},\ }\bibfield  {title} {\enquote {\bibinfo
  {title} {{Symmetry TFTs from String Theory}},}\ }\href {\doibase
  10.1007/s00220-023-04737-2} {\bibfield  {journal} {\bibinfo  {journal}
  {Commun. Math. Phys.}\ }\textbf {\bibinfo {volume} {402}},\ \bibinfo {pages}
  {895--949} (\bibinfo {year} {2023})},\ \Eprint
  {http://arxiv.org/abs/2112.02092} {arXiv:2112.02092 [hep-th]} \BibitemShut
  {NoStop}%
\bibitem [{\citenamefont {Chatterjee}\ and\ \citenamefont
  {Wen}(2023{\natexlab{b}})}]{Chatterjee:2022kxb}%
  \BibitemOpen
  \bibfield  {author} {\bibinfo {author} {\bibfnamefont {Arkya}\ \bibnamefont
  {Chatterjee}}\ and\ \bibinfo {author} {\bibfnamefont {Xiao-Gang}\
  \bibnamefont {Wen}},\ }\bibfield  {title} {\enquote {\bibinfo {title}
  {{Symmetry as a shadow of topological order and a derivation of topological
  holographic principle}},}\ }\href {\doibase 10.1103/PhysRevB.107.155136}
  {\bibfield  {journal} {\bibinfo  {journal} {Phys. Rev. B}\ }\textbf {\bibinfo
  {volume} {107}},\ \bibinfo {pages} {155136} (\bibinfo {year}
  {2023}{\natexlab{b}})},\ \Eprint {http://arxiv.org/abs/2203.03596}
  {arXiv:2203.03596 [cond-mat.str-el]} \BibitemShut {NoStop}%
\bibitem [{\citenamefont {Freed}\ and\ \citenamefont
  {Teleman}(2022)}]{Freed:2018cec}%
  \BibitemOpen
  \bibfield  {author} {\bibinfo {author} {\bibfnamefont {Daniel~S.}\
  \bibnamefont {Freed}}\ and\ \bibinfo {author} {\bibfnamefont {Constantin}\
  \bibnamefont {Teleman}},\ }\bibfield  {title} {\enquote {\bibinfo {title}
  {{Topological dualities in the Ising model}},}\ }\href {\doibase
  10.2140/gt.2022.26.1907} {\bibfield  {journal} {\bibinfo  {journal} {Geom.
  Topol.}\ }\textbf {\bibinfo {volume} {26}},\ \bibinfo {pages} {1907--1984}
  (\bibinfo {year} {2022})},\ \Eprint {http://arxiv.org/abs/1806.00008}
  {arXiv:1806.00008 [math.AT]} \BibitemShut {NoStop}%
\bibitem [{\citenamefont {Kong}\ and\ \citenamefont
  {Zheng}(2020)}]{Kong:2019byq}%
  \BibitemOpen
  \bibfield  {author} {\bibinfo {author} {\bibfnamefont {Liang}\ \bibnamefont
  {Kong}}\ and\ \bibinfo {author} {\bibfnamefont {Hao}\ \bibnamefont {Zheng}},\
  }\bibfield  {title} {\enquote {\bibinfo {title} {{A mathematical theory of
  gapless edges of 2d topological orders. Part I}},}\ }\href {\doibase
  10.1007/JHEP02(2020)150} {\bibfield  {journal} {\bibinfo  {journal} {JHEP}\
  }\textbf {\bibinfo {volume} {02}},\ \bibinfo {pages} {150} (\bibinfo {year}
  {2020})},\ \Eprint {http://arxiv.org/abs/1905.04924} {arXiv:1905.04924
  [cond-mat.str-el]} \BibitemShut {NoStop}%
\bibitem [{\citenamefont {Bhardwaj}\ and\ \citenamefont
  {Tachikawa}(2018)}]{Bhardwaj:2017xup}%
  \BibitemOpen
  \bibfield  {author} {\bibinfo {author} {\bibfnamefont {Lakshya}\ \bibnamefont
  {Bhardwaj}}\ and\ \bibinfo {author} {\bibfnamefont {Yuji}\ \bibnamefont
  {Tachikawa}},\ }\bibfield  {title} {\enquote {\bibinfo {title} {{On finite
  symmetries and their gauging in two dimensions}},}\ }\href {\doibase
  10.1007/JHEP03(2018)189} {\bibfield  {journal} {\bibinfo  {journal} {JHEP}\
  }\textbf {\bibinfo {volume} {03}},\ \bibinfo {pages} {189} (\bibinfo {year}
  {2018})},\ \Eprint {http://arxiv.org/abs/1704.02330} {arXiv:1704.02330
  [hep-th]} \BibitemShut {NoStop}%
\bibitem [{\citenamefont {Albert}\ \emph {et~al.}(2021)\citenamefont {Albert},
  \citenamefont {Aasen}, \citenamefont {Xu}, \citenamefont {Ji}, \citenamefont
  {Alicea},\ and\ \citenamefont {Preskill}}]{Albert:2021vts}%
  \BibitemOpen
  \bibfield  {author} {\bibinfo {author} {\bibfnamefont {Victor~V.}\
  \bibnamefont {Albert}}, \bibinfo {author} {\bibfnamefont {David}\
  \bibnamefont {Aasen}}, \bibinfo {author} {\bibfnamefont {Wenqing}\
  \bibnamefont {Xu}}, \bibinfo {author} {\bibfnamefont {Wenjie}\ \bibnamefont
  {Ji}}, \bibinfo {author} {\bibfnamefont {Jason}\ \bibnamefont {Alicea}}, \
  and\ \bibinfo {author} {\bibfnamefont {John}\ \bibnamefont {Preskill}},\
  }\bibfield  {title} {\enquote {\bibinfo {title} {{Spin chains, defects, and
  quantum wires for the quantum-double edge}},}\ }\href@noop {} {\  (\bibinfo
  {year} {2021})},\ \Eprint {http://arxiv.org/abs/2111.12096} {arXiv:2111.12096
  [cond-mat.str-el]} \BibitemShut {NoStop}%
\bibitem [{\citenamefont {Vanhove}\ \emph {et~al.}(2018)\citenamefont
  {Vanhove}, \citenamefont {Bal}, \citenamefont {Williamson}, \citenamefont
  {Bultinck}, \citenamefont {Haegeman},\ and\ \citenamefont
  {Verstraete}}]{Vanhove:2018wlb}%
  \BibitemOpen
  \bibfield  {author} {\bibinfo {author} {\bibfnamefont {Robijn}\ \bibnamefont
  {Vanhove}}, \bibinfo {author} {\bibfnamefont {Matthias}\ \bibnamefont {Bal}},
  \bibinfo {author} {\bibfnamefont {Dominic~J.}\ \bibnamefont {Williamson}},
  \bibinfo {author} {\bibfnamefont {Nick}\ \bibnamefont {Bultinck}}, \bibinfo
  {author} {\bibfnamefont {Jutho}\ \bibnamefont {Haegeman}}, \ and\ \bibinfo
  {author} {\bibfnamefont {Frank}\ \bibnamefont {Verstraete}},\ }\bibfield
  {title} {\enquote {\bibinfo {title} {{Mapping topological to conformal field
  theories through strange correlators}},}\ }\href {\doibase
  10.1103/PhysRevLett.121.177203} {\bibfield  {journal} {\bibinfo  {journal}
  {Phys. Rev. Lett.}\ }\textbf {\bibinfo {volume} {121}},\ \bibinfo {pages}
  {177203} (\bibinfo {year} {2018})},\ \Eprint
  {http://arxiv.org/abs/1801.05959} {arXiv:1801.05959 [quant-ph]} \BibitemShut
  {NoStop}%
\bibitem [{\citenamefont {Aasen}\ \emph {et~al.}(2020)\citenamefont {Aasen},
  \citenamefont {Fendley},\ and\ \citenamefont {Mong}}]{Aasen:2020jwb}%
  \BibitemOpen
  \bibfield  {author} {\bibinfo {author} {\bibfnamefont {David}\ \bibnamefont
  {Aasen}}, \bibinfo {author} {\bibfnamefont {Paul}\ \bibnamefont {Fendley}}, \
  and\ \bibinfo {author} {\bibfnamefont {Roger S.~K.}\ \bibnamefont {Mong}},\
  }\bibfield  {title} {\enquote {\bibinfo {title} {{Topological Defects on the
  Lattice: Dualities and Degeneracies}},}\ }\href@noop {} {\  (\bibinfo {year}
  {2020})},\ \Eprint {http://arxiv.org/abs/2008.08598} {arXiv:2008.08598
  [cond-mat.stat-mech]} \BibitemShut {NoStop}%
\bibitem [{\citenamefont {Chen}\ \emph {et~al.}(2024)\citenamefont {Chen},
  \citenamefont {Ji}, \citenamefont {Zhang}, \citenamefont {Shen},
  \citenamefont {Wang}, \citenamefont {Zeng},\ and\ \citenamefont
  {Hung}}]{Chen:2022wvy}%
  \BibitemOpen
  \bibfield  {author} {\bibinfo {author} {\bibfnamefont {Lin}\ \bibnamefont
  {Chen}}, \bibinfo {author} {\bibfnamefont {Kaixin}\ \bibnamefont {Ji}},
  \bibinfo {author} {\bibfnamefont {Haochen}\ \bibnamefont {Zhang}}, \bibinfo
  {author} {\bibfnamefont {Ce}~\bibnamefont {Shen}}, \bibinfo {author}
  {\bibfnamefont {Ruoshui}\ \bibnamefont {Wang}}, \bibinfo {author}
  {\bibfnamefont {Xiangdong}\ \bibnamefont {Zeng}}, \ and\ \bibinfo {author}
  {\bibfnamefont {Ling-Yan}\ \bibnamefont {Hung}},\ }\bibfield  {title}
  {\enquote {\bibinfo {title} {{CFTD from TQFTD+1 via Holographic Tensor
  Network, and Precision Discretization of CFT2 }},}\ }\href {\doibase
  10.1103/PhysRevX.14.041033} {\bibfield  {journal} {\bibinfo  {journal} {Phys.
  Rev. X}\ }\textbf {\bibinfo {volume} {14}},\ \bibinfo {pages} {041033}
  (\bibinfo {year} {2024})},\ \Eprint {http://arxiv.org/abs/2210.12127}
  {arXiv:2210.12127 [hep-th]} \BibitemShut {NoStop}%
\bibitem [{\citenamefont {Levin}\ and\ \citenamefont
  {Wen}(2005)}]{Levin:2004mi}%
  \BibitemOpen
  \bibfield  {author} {\bibinfo {author} {\bibfnamefont {Michael~A.}\
  \bibnamefont {Levin}}\ and\ \bibinfo {author} {\bibfnamefont {Xiao-Gang}\
  \bibnamefont {Wen}},\ }\bibfield  {title} {\enquote {\bibinfo {title}
  {{String net condensation: A Physical mechanism for topological phases}},}\
  }\href {\doibase 10.1103/PhysRevB.71.045110} {\bibfield  {journal} {\bibinfo
  {journal} {Phys. Rev. B}\ }\textbf {\bibinfo {volume} {71}},\ \bibinfo
  {pages} {045110} (\bibinfo {year} {2005})},\ \Eprint
  {http://arxiv.org/abs/cond-mat/0404617} {arXiv:cond-mat/0404617} \BibitemShut
  {NoStop}%
\bibitem [{\citenamefont {Turaev}\ and\ \citenamefont
  {Viro}(1992)}]{Turaev:1992hq}%
  \BibitemOpen
  \bibfield  {author} {\bibinfo {author} {\bibfnamefont {V.~G.}\ \bibnamefont
  {Turaev}}\ and\ \bibinfo {author} {\bibfnamefont {O.~Y.}\ \bibnamefont
  {Viro}},\ }\bibfield  {title} {\enquote {\bibinfo {title} {{State sum
  invariants of 3 manifolds and quantum 6j symbols}},}\ }\href {\doibase
  10.1016/0040-9383(92)90015-A} {\bibfield  {journal} {\bibinfo  {journal}
  {Topology}\ }\textbf {\bibinfo {volume} {31}},\ \bibinfo {pages} {865--902}
  (\bibinfo {year} {1992})}\BibitemShut {NoStop}%
\bibitem [{\citenamefont {Gaiotto}\ \emph {et~al.}(2015)\citenamefont
  {Gaiotto}, \citenamefont {Kapustin}, \citenamefont {Seiberg},\ and\
  \citenamefont {Willett}}]{Gaiotto:2014kfa}%
  \BibitemOpen
  \bibfield  {author} {\bibinfo {author} {\bibfnamefont {Davide}\ \bibnamefont
  {Gaiotto}}, \bibinfo {author} {\bibfnamefont {Anton}\ \bibnamefont
  {Kapustin}}, \bibinfo {author} {\bibfnamefont {Nathan}\ \bibnamefont
  {Seiberg}}, \ and\ \bibinfo {author} {\bibfnamefont {Brian}\ \bibnamefont
  {Willett}},\ }\bibfield  {title} {\enquote {\bibinfo {title} {{Generalized
  Global Symmetries}},}\ }\href {\doibase 10.1007/JHEP02(2015)172} {\bibfield
  {journal} {\bibinfo  {journal} {JHEP}\ }\textbf {\bibinfo {volume} {02}},\
  \bibinfo {pages} {172} (\bibinfo {year} {2015})},\ \Eprint
  {http://arxiv.org/abs/1412.5148} {arXiv:1412.5148 [hep-th]} \BibitemShut
  {NoStop}%
\bibitem [{\citenamefont {Moore}\ and\ \citenamefont
  {Seiberg}(1989)}]{Moore:1988qv}%
  \BibitemOpen
  \bibfield  {author} {\bibinfo {author} {\bibfnamefont {Gregory~W.}\
  \bibnamefont {Moore}}\ and\ \bibinfo {author} {\bibfnamefont {Nathan}\
  \bibnamefont {Seiberg}},\ }\bibfield  {title} {\enquote {\bibinfo {title}
  {{Classical and Quantum Conformal Field Theory}},}\ }\href {\doibase
  10.1007/BF01238857} {\bibfield  {journal} {\bibinfo  {journal} {Commun. Math.
  Phys.}\ }\textbf {\bibinfo {volume} {123}},\ \bibinfo {pages} {177} (\bibinfo
  {year} {1989})}\BibitemShut {NoStop}%
\bibitem [{\citenamefont {Verlinde}(1988)}]{Verlinde:1988sn}%
  \BibitemOpen
  \bibfield  {author} {\bibinfo {author} {\bibfnamefont {Erik~P.}\ \bibnamefont
  {Verlinde}},\ }\bibfield  {title} {\enquote {\bibinfo {title} {{Fusion Rules
  and Modular Transformations in 2D Conformal Field Theory}},}\ }\href
  {\doibase 10.1016/0550-3213(88)90603-7} {\bibfield  {journal} {\bibinfo
  {journal} {Nucl. Phys. B}\ }\textbf {\bibinfo {volume} {300}},\ \bibinfo
  {pages} {360--376} (\bibinfo {year} {1988})}\BibitemShut {NoStop}%
\bibitem [{\citenamefont {Petkova}\ and\ \citenamefont
  {Zuber}(2001)}]{Petkova:2000ip}%
  \BibitemOpen
  \bibfield  {author} {\bibinfo {author} {\bibfnamefont {V.~B.}\ \bibnamefont
  {Petkova}}\ and\ \bibinfo {author} {\bibfnamefont {J.~B.}\ \bibnamefont
  {Zuber}},\ }\bibfield  {title} {\enquote {\bibinfo {title} {{Generalized
  twisted partition functions}},}\ }\href {\doibase
  10.1016/S0370-2693(01)00276-3} {\bibfield  {journal} {\bibinfo  {journal}
  {Phys. Lett. B}\ }\textbf {\bibinfo {volume} {504}},\ \bibinfo {pages}
  {157--164} (\bibinfo {year} {2001})},\ \Eprint
  {http://arxiv.org/abs/hep-th/0011021} {arXiv:hep-th/0011021} \BibitemShut
  {NoStop}%
\bibitem [{\citenamefont {Fuchs}\ \emph {et~al.}(2002)\citenamefont {Fuchs},
  \citenamefont {Runkel},\ and\ \citenamefont {Schweigert}}]{Fuchs:2002cm}%
  \BibitemOpen
  \bibfield  {author} {\bibinfo {author} {\bibfnamefont {Jurgen}\ \bibnamefont
  {Fuchs}}, \bibinfo {author} {\bibfnamefont {Ingo}\ \bibnamefont {Runkel}}, \
  and\ \bibinfo {author} {\bibfnamefont {Christoph}\ \bibnamefont
  {Schweigert}},\ }\bibfield  {title} {\enquote {\bibinfo {title} {{TFT
  construction of RCFT correlators 1. Partition functions}},}\ }\href {\doibase
  10.1016/S0550-3213(02)00744-7} {\bibfield  {journal} {\bibinfo  {journal}
  {Nucl. Phys. B}\ }\textbf {\bibinfo {volume} {646}},\ \bibinfo {pages}
  {353--497} (\bibinfo {year} {2002})},\ \Eprint
  {http://arxiv.org/abs/hep-th/0204148} {arXiv:hep-th/0204148} \BibitemShut
  {NoStop}%
\bibitem [{\citenamefont {Frohlich}\ \emph {et~al.}(2007)\citenamefont
  {Frohlich}, \citenamefont {Fuchs}, \citenamefont {Runkel},\ and\
  \citenamefont {Schweigert}}]{Frohlich:2006ch}%
  \BibitemOpen
  \bibfield  {author} {\bibinfo {author} {\bibfnamefont {Jurg}\ \bibnamefont
  {Frohlich}}, \bibinfo {author} {\bibfnamefont {Jurgen}\ \bibnamefont
  {Fuchs}}, \bibinfo {author} {\bibfnamefont {Ingo}\ \bibnamefont {Runkel}}, \
  and\ \bibinfo {author} {\bibfnamefont {Christoph}\ \bibnamefont
  {Schweigert}},\ }\bibfield  {title} {\enquote {\bibinfo {title} {{Duality and
  defects in rational conformal field theory}},}\ }\href {\doibase
  10.1016/j.nuclphysb.2006.11.017} {\bibfield  {journal} {\bibinfo  {journal}
  {Nucl. Phys. B}\ }\textbf {\bibinfo {volume} {763}},\ \bibinfo {pages}
  {354--430} (\bibinfo {year} {2007})},\ \Eprint
  {http://arxiv.org/abs/hep-th/0607247} {arXiv:hep-th/0607247} \BibitemShut
  {NoStop}%
\bibitem [{\citenamefont {Chang}\ \emph {et~al.}(2019)\citenamefont {Chang},
  \citenamefont {Lin}, \citenamefont {Shao}, \citenamefont {Wang},\ and\
  \citenamefont {Yin}}]{Chang:2018iay}%
  \BibitemOpen
  \bibfield  {author} {\bibinfo {author} {\bibfnamefont {Chi-Ming}\
  \bibnamefont {Chang}}, \bibinfo {author} {\bibfnamefont {Ying-Hsuan}\
  \bibnamefont {Lin}}, \bibinfo {author} {\bibfnamefont {Shu-Heng}\
  \bibnamefont {Shao}}, \bibinfo {author} {\bibfnamefont {Yifan}\ \bibnamefont
  {Wang}}, \ and\ \bibinfo {author} {\bibfnamefont {Xi}~\bibnamefont {Yin}},\
  }\bibfield  {title} {\enquote {\bibinfo {title} {{Topological Defect Lines
  and Renormalization Group Flows in Two Dimensions}},}\ }\href {\doibase
  10.1007/JHEP01(2019)026} {\bibfield  {journal} {\bibinfo  {journal} {JHEP}\
  }\textbf {\bibinfo {volume} {01}},\ \bibinfo {pages} {026} (\bibinfo {year}
  {2019})},\ \Eprint {http://arxiv.org/abs/1802.04445} {arXiv:1802.04445
  [hep-th]} \BibitemShut {NoStop}%
\bibitem [{\citenamefont {Thorngren}(2020)}]{Thorngren:2018bhj}%
  \BibitemOpen
  \bibfield  {author} {\bibinfo {author} {\bibfnamefont {Ryan}\ \bibnamefont
  {Thorngren}},\ }\bibfield  {title} {\enquote {\bibinfo {title} {{Anomalies
  and Bosonization}},}\ }\href {\doibase 10.1007/s00220-020-03830-0} {\bibfield
   {journal} {\bibinfo  {journal} {Commun. Math. Phys.}\ }\textbf {\bibinfo
  {volume} {378}},\ \bibinfo {pages} {1775--1816} (\bibinfo {year} {2020})},\
  \Eprint {http://arxiv.org/abs/1810.04414} {arXiv:1810.04414
  [cond-mat.str-el]} \BibitemShut {NoStop}%
\bibitem [{\citenamefont {Cardy}(1984)}]{CARDY1984514}%
  \BibitemOpen
  \bibfield  {author} {\bibinfo {author} {\bibfnamefont {John~L.}\ \bibnamefont
  {Cardy}},\ }\bibfield  {title} {\enquote {\bibinfo {title} {Conformal
  invariance and surface critical behavior},}\ }\href {\doibase
  https://doi.org/10.1016/0550-3213(84)90241-4} {\bibfield  {journal} {\bibinfo
   {journal} {Nuclear Physics B}\ }\textbf {\bibinfo {volume} {240}},\ \bibinfo
  {pages} {514--532} (\bibinfo {year} {1984})}\BibitemShut {NoStop}%
\bibitem [{\citenamefont {Cardy}(1989)}]{CARDY1989581}%
  \BibitemOpen
  \bibfield  {author} {\bibinfo {author} {\bibfnamefont {John~L.}\ \bibnamefont
  {Cardy}},\ }\bibfield  {title} {\enquote {\bibinfo {title} {Boundary
  conditions, fusion rules and the verlinde formula},}\ }\href {\doibase
  https://doi.org/10.1016/0550-3213(89)90521-X} {\bibfield  {journal} {\bibinfo
   {journal} {Nuclear Physics B}\ }\textbf {\bibinfo {volume} {324}},\ \bibinfo
  {pages} {581--596} (\bibinfo {year} {1989})}\BibitemShut {NoStop}%
\bibitem [{\citenamefont {ISHIBASHI}(1989)}]{doi:10.1142/S0217732389000320}%
  \BibitemOpen
  \bibfield  {author} {\bibinfo {author} {\bibfnamefont {NOBUYUKI}\
  \bibnamefont {ISHIBASHI}},\ }\bibfield  {title} {\enquote {\bibinfo {title}
  {The boundary and crosscap states in conformal field theories},}\ }\href
  {\doibase 10.1142/S0217732389000320} {\bibfield  {journal} {\bibinfo
  {journal} {Modern Physics Letters A}\ }\textbf {\bibinfo {volume} {04}},\
  \bibinfo {pages} {251--264} (\bibinfo {year} {1989})},\ \Eprint
  {http://arxiv.org/abs/https://doi.org/10.1142/S0217732389000320}
  {https://doi.org/10.1142/S0217732389000320} \BibitemShut {NoStop}%
\bibitem [{\citenamefont {ONOGI}\ and\ \citenamefont
  {ISHIBASHI}(1989)}]{doi:10.1142/S0217732389000228}%
  \BibitemOpen
  \bibfield  {author} {\bibinfo {author} {\bibfnamefont {TETSUYA}\ \bibnamefont
  {ONOGI}}\ and\ \bibinfo {author} {\bibfnamefont {NOBUYUKI}\ \bibnamefont
  {ISHIBASHI}},\ }\bibfield  {title} {\enquote {\bibinfo {title} {Conformal
  field theories on surfaces with boundaries and crosscaps},}\ }\href {\doibase
  10.1142/S0217732389000228} {\bibfield  {journal} {\bibinfo  {journal} {Modern
  Physics Letters A}\ }\textbf {\bibinfo {volume} {04}},\ \bibinfo {pages}
  {161--168} (\bibinfo {year} {1989})},\ \Eprint
  {http://arxiv.org/abs/https://doi.org/10.1142/S0217732389000228}
  {https://doi.org/10.1142/S0217732389000228} \BibitemShut {NoStop}%
\bibitem [{\citenamefont {Cappelli}\ \emph {et~al.}(1987)\citenamefont
  {Cappelli}, \citenamefont {Itzykson},\ and\ \citenamefont
  {Zuber}}]{CAPPELLI1987445}%
  \BibitemOpen
  \bibfield  {author} {\bibinfo {author} {\bibfnamefont {A.}~\bibnamefont
  {Cappelli}}, \bibinfo {author} {\bibfnamefont {C.}~\bibnamefont {Itzykson}},
  \ and\ \bibinfo {author} {\bibfnamefont {J.-B.}\ \bibnamefont {Zuber}},\
  }\bibfield  {title} {\enquote {\bibinfo {title} {Modular invariant partition
  functions in two dimensions},}\ }\href {\doibase
  https://doi.org/10.1016/0550-3213(87)90155-6} {\bibfield  {journal} {\bibinfo
   {journal} {Nuclear Physics B}\ }\textbf {\bibinfo {volume} {280}},\ \bibinfo
  {pages} {445--465} (\bibinfo {year} {1987})}\BibitemShut {NoStop}%
\bibitem [{\citenamefont {Cardy}(2008)}]{Cardy:2008jc}%
  \BibitemOpen
  \bibfield  {author} {\bibinfo {author} {\bibfnamefont {John}\ \bibnamefont
  {Cardy}},\ }\bibfield  {title} {\enquote {\bibinfo {title} {{Conformal Field
  Theory and Statistical Mechanics}},}\ }in\ \href@noop {} {\emph {\bibinfo
  {booktitle} {{Les Houches Summer School: Session 89: Exacts Methods in
  Low-Dimensional Statistical Physics and Quantum Computing}}}}\ (\bibinfo
  {year} {2008})\ \Eprint {http://arxiv.org/abs/0807.3472} {arXiv:0807.3472
  [cond-mat.stat-mech]} \BibitemShut {NoStop}%
\bibitem [{\citenamefont {Di~Francesco}\ \emph {et~al.}(1997)\citenamefont
  {Di~Francesco}, \citenamefont {Mathieu},\ and\ \citenamefont
  {Senechal}}]{DiFrancesco:1997nk}%
  \BibitemOpen
  \bibfield  {author} {\bibinfo {author} {\bibfnamefont {P.}~\bibnamefont
  {Di~Francesco}}, \bibinfo {author} {\bibfnamefont {P.}~\bibnamefont
  {Mathieu}}, \ and\ \bibinfo {author} {\bibfnamefont {D.}~\bibnamefont
  {Senechal}},\ }\href {\doibase 10.1007/978-1-4612-2256-9} {\emph {\bibinfo
  {title} {{Conformal Field Theory}}}},\ Graduate Texts in Contemporary
  Physics\ (\bibinfo  {publisher} {Springer-Verlag},\ \bibinfo {address} {New
  York},\ \bibinfo {year} {1997})\BibitemShut {NoStop}%
\bibitem [{\citenamefont {Cardy}\ and\ \citenamefont
  {Lewellen}(1991)}]{CARDY1991274}%
  \BibitemOpen
  \bibfield  {author} {\bibinfo {author} {\bibfnamefont {John~L.}\ \bibnamefont
  {Cardy}}\ and\ \bibinfo {author} {\bibfnamefont {David~C.}\ \bibnamefont
  {Lewellen}},\ }\bibfield  {title} {\enquote {\bibinfo {title} {Bulk and
  boundary operators in conformal field theory},}\ }\href {\doibase
  https://doi.org/10.1016/0370-2693(91)90828-E} {\bibfield  {journal} {\bibinfo
   {journal} {Physics Letters B}\ }\textbf {\bibinfo {volume} {259}},\ \bibinfo
  {pages} {274--278} (\bibinfo {year} {1991})}\BibitemShut {NoStop}%
\bibitem [{\citenamefont {Lewellen}(1992)}]{LEWELLEN1992654}%
  \BibitemOpen
  \bibfield  {author} {\bibinfo {author} {\bibfnamefont {David~C.}\
  \bibnamefont {Lewellen}},\ }\bibfield  {title} {\enquote {\bibinfo {title}
  {Sewing constraints for conformal field theories on surfaces with
  boundaries},}\ }\href {\doibase https://doi.org/10.1016/0550-3213(92)90370-Q}
  {\bibfield  {journal} {\bibinfo  {journal} {Nuclear Physics B}\ }\textbf
  {\bibinfo {volume} {372}},\ \bibinfo {pages} {654--682} (\bibinfo {year}
  {1992})}\BibitemShut {NoStop}%
\bibitem [{\citenamefont {Kojita}\ \emph {et~al.}(2018)\citenamefont {Kojita},
  \citenamefont {Maccaferri}, \citenamefont {Masuda},\ and\ \citenamefont
  {Schnabl}}]{Kojita:2016jwe}%
  \BibitemOpen
  \bibfield  {author} {\bibinfo {author} {\bibfnamefont {Toshiko}\ \bibnamefont
  {Kojita}}, \bibinfo {author} {\bibfnamefont {Carlo}\ \bibnamefont
  {Maccaferri}}, \bibinfo {author} {\bibfnamefont {Toru}\ \bibnamefont
  {Masuda}}, \ and\ \bibinfo {author} {\bibfnamefont {Martin}\ \bibnamefont
  {Schnabl}},\ }\bibfield  {title} {\enquote {\bibinfo {title} {{Topological
  defects in open string field theory}},}\ }\href {\doibase
  10.1007/JHEP04(2018)057} {\bibfield  {journal} {\bibinfo  {journal} {JHEP}\
  }\textbf {\bibinfo {volume} {04}},\ \bibinfo {pages} {057} (\bibinfo {year}
  {2018})},\ \Eprint {http://arxiv.org/abs/1612.01997} {arXiv:1612.01997
  [hep-th]} \BibitemShut {NoStop}%
\bibitem [{\citenamefont {Fuchs}\ \emph {et~al.}(2005)\citenamefont {Fuchs},
  \citenamefont {Runkel},\ and\ \citenamefont {Schweigert}}]{Fuchs:2004xi}%
  \BibitemOpen
  \bibfield  {author} {\bibinfo {author} {\bibfnamefont {Jurgen}\ \bibnamefont
  {Fuchs}}, \bibinfo {author} {\bibfnamefont {Ingo}\ \bibnamefont {Runkel}}, \
  and\ \bibinfo {author} {\bibfnamefont {Christoph}\ \bibnamefont
  {Schweigert}},\ }\bibfield  {title} {\enquote {\bibinfo {title} {{TFT
  construction of RCFT correlators IV: Structure constants and correlation
  functions}},}\ }\href {\doibase 10.1016/j.nuclphysb.2005.03.018} {\bibfield
  {journal} {\bibinfo  {journal} {Nucl. Phys. B}\ }\textbf {\bibinfo {volume}
  {715}},\ \bibinfo {pages} {539--638} (\bibinfo {year} {2005})},\ \Eprint
  {http://arxiv.org/abs/hep-th/0412290} {arXiv:hep-th/0412290} \BibitemShut
  {NoStop}%
\bibitem [{\citenamefont {Gu}\ \emph {et~al.}(2009)\citenamefont {Gu},
  \citenamefont {Levin}, \citenamefont {Swingle},\ and\ \citenamefont
  {Wen}}]{wavefunction1}%
  \BibitemOpen
  \bibfield  {author} {\bibinfo {author} {\bibfnamefont {Zheng-Cheng}\
  \bibnamefont {Gu}}, \bibinfo {author} {\bibfnamefont {Michael}\ \bibnamefont
  {Levin}}, \bibinfo {author} {\bibfnamefont {Brian}\ \bibnamefont {Swingle}},
  \ and\ \bibinfo {author} {\bibfnamefont {Xiao-Gang}\ \bibnamefont {Wen}},\
  }\bibfield  {title} {\enquote {\bibinfo {title} {Tensor-product
  representations for string-net condensed states},}\ }\href {\doibase
  10.1103/PhysRevB.79.085118} {\bibfield  {journal} {\bibinfo  {journal} {Phys.
  Rev. B}\ }\textbf {\bibinfo {volume} {79}},\ \bibinfo {pages} {085118}
  (\bibinfo {year} {2009})}\BibitemShut {NoStop}%
\bibitem [{\citenamefont {Buerschaper}\ \emph {et~al.}(2009)\citenamefont
  {Buerschaper}, \citenamefont {Aguado},\ and\ \citenamefont
  {Vidal}}]{wavefunction2}%
  \BibitemOpen
  \bibfield  {author} {\bibinfo {author} {\bibfnamefont {Oliver}\ \bibnamefont
  {Buerschaper}}, \bibinfo {author} {\bibfnamefont {Miguel}\ \bibnamefont
  {Aguado}}, \ and\ \bibinfo {author} {\bibfnamefont {Guifr\'e}\ \bibnamefont
  {Vidal}},\ }\bibfield  {title} {\enquote {\bibinfo {title} {Explicit tensor
  network representation for the ground states of string-net models},}\ }\href
  {\doibase 10.1103/PhysRevB.79.085119} {\bibfield  {journal} {\bibinfo
  {journal} {Phys. Rev. B}\ }\textbf {\bibinfo {volume} {79}},\ \bibinfo
  {pages} {085119} (\bibinfo {year} {2009})}\BibitemShut {NoStop}%
\bibitem [{\citenamefont {Kirillov}\ and\ \citenamefont
  {Reshetikhin}(1991)}]{Kirillov:1991ec}%
  \BibitemOpen
  \bibfield  {author} {\bibinfo {author} {\bibfnamefont {A.~N.}\ \bibnamefont
  {Kirillov}}\ and\ \bibinfo {author} {\bibfnamefont {N.~Yu.}\ \bibnamefont
  {Reshetikhin}},\ }\bibfield  {title} {\enquote {\bibinfo {title}
  {{Representations of the algebra $U_q(sl(2))$, q-orthogonal polynomials and
  invariants of links}},}\ }\href
  {https://math.berkeley.edu/~reshetik/Publications/q6j-KR.pdf} {\bibfield
  {journal} {\bibinfo  {journal} {New developments in the theory of knots}\ ,\
  \bibinfo {pages} {202--256}} (\bibinfo {year} {1991})}\BibitemShut {NoStop}%
\bibitem [{\citenamefont {Dotsenko}\ and\ \citenamefont
  {Fateev}(1985)}]{DOTSENKO1985291}%
  \BibitemOpen
  \bibfield  {author} {\bibinfo {author} {\bibfnamefont {Vl.S.}\ \bibnamefont
  {Dotsenko}}\ and\ \bibinfo {author} {\bibfnamefont {V.A.}\ \bibnamefont
  {Fateev}},\ }\bibfield  {title} {\enquote {\bibinfo {title} {Operator algebra
  of two-dimensional conformal theories with central charge $c \leq 1$},}\
  }\href {\doibase https://doi.org/10.1016/0370-2693(85)90366-1} {\bibfield
  {journal} {\bibinfo  {journal} {Physics Letters B}\ }\textbf {\bibinfo
  {volume} {154}},\ \bibinfo {pages} {291--295} (\bibinfo {year}
  {1985})}\BibitemShut {NoStop}%
\bibitem [{\citenamefont {Bal}\ \emph {et~al.}(2017)\citenamefont {Bal},
  \citenamefont {Mari\"en}, \citenamefont {Haegeman},\ and\ \citenamefont
  {Verstraete}}]{PhysRevLett.118.250602}%
  \BibitemOpen
  \bibfield  {author} {\bibinfo {author} {\bibfnamefont {M.}~\bibnamefont
  {Bal}}, \bibinfo {author} {\bibfnamefont {M.}~\bibnamefont {Mari\"en}},
  \bibinfo {author} {\bibfnamefont {J.}~\bibnamefont {Haegeman}}, \ and\
  \bibinfo {author} {\bibfnamefont {F.}~\bibnamefont {Verstraete}},\ }\bibfield
   {title} {\enquote {\bibinfo {title} {Renormalization group flows of
  hamiltonians using tensor networks},}\ }\href {\doibase
  10.1103/PhysRevLett.118.250602} {\bibfield  {journal} {\bibinfo  {journal}
  {Phys. Rev. Lett.}\ }\textbf {\bibinfo {volume} {118}},\ \bibinfo {pages}
  {250602} (\bibinfo {year} {2017})}\BibitemShut {NoStop}%
\end{thebibliography}%

\end{document}